\documentclass[12pt]{article}
\usepackage{amsmath,bbm}
\usepackage{amssymb}
\usepackage{graphicx,psfrag,epsf}
\usepackage{enumerate}
\usepackage{natbib}
\usepackage{caption}
\usepackage{mathtools}

\usepackage{subcaption}
\newcommand{\PPPlotFrameB}[1]{%
\includegraphics[width=\textwidth]{#1}\endgroup}
\usepackage{xcolor}

\def\PPPlotFrame{\begingroup 
\catcode`\_=12
\PPPlotFrameB}

\newcommand{\blind}{0}

\makeatletter
\newcommand{\blackub}{}
\def\blackub#1{%
  \@ifnextchar_%
    {\@redub{#1}}
    {\@latex@warning{Missing argument for \string\blackub}\@redub{#1}_{}}%
}
\def\@redub#1_#2{%
    \colorlet{currentcolor}{.}%
    \color{black}%
    \underbrace{\color{currentcolor}#1}_{\color{black}#2}%
    \color{currentcolor}%
}
\makeatother

\begin{document}

\def\spacingset#1{\renewcommand{\baselinestretch}%
{#1}\small\normalsize} \spacingset{1}


\if0\blind
{
  \title{\bf Bayesian outcome-guided multi-view mixture models with applications in molecular precision medicine}
  \author{Paul D. W. Kirk\\
    and \\
    Filippo Pagani \\
    and \\
    Sylvia Richardson \\
    MRC Biostatistics Unit, Cambridge, UK}
  \maketitle
} \fi

\if1\blind
{
  \bigskip
  \bigskip
  \bigskip
  \begin{center}
    {\LARGE\bf Semi-supervised multi-view clustering}
\end{center}
  \medskip
} \fi

\bigskip
\begin{abstract}
Clustering is commonly performed as an initial analysis step for uncovering structure in 'omics datasets, e.g. to discover molecular subtypes of disease.  The high-throughput, high-dimensional nature of these datasets means that they provide information on a diverse array of different biomolecular processes and pathways.  Different groups of variables (e.g. genes or proteins) will be implicated in different biomolecular processes, and hence undertaking analyses that are limited to identifying just a single clustering partition of the whole dataset is therefore liable to conflate the {\em multiple} clustering structures that may arise from these distinct processes.  To address this, we propose a multi-view Bayesian mixture model that identifies groups of variables (``views"), each of which defines a distinct clustering structure.  We consider applications in stratified medicine, for which our principal goal is to identify clusters of patients that define distinct, clinically actionable disease subtypes.  We adopt the semi-supervised, outcome-guided mixture modelling approach of {\em Bayesian profile regression} that makes use of a response variable in order
to guide inference toward the clusterings that are most relevant in a stratified medicine context.  We present the model, together with
illustrative simulation examples, and examples from pan-cancer proteomics. We demonstrate how the approach can be used to perform integrative clustering, and consider an example in which different 'omics datasets are integrated in the context of breast cancer subtyping. 
\end{abstract}

\noindent%

\spacingset{1.45}
\section{Introduction}
\label{sec:intro}

Clustering is ubiquoitously used in the analysis of omics data as a means to uncover structure and patterns in these large, high-dimensional datasets \citep[e.g.][]{Eisen1998,Heyer:1999,Alon:1999,BenDor:1999,Son:2005}.  Here we are particularly interested in molecular precision medicine applications, in which the analysis objective is to identify molecular subtypes of disease \citep[e.g.][]{Golub1999,Perou2000,Sorlie2001,Cancer::2012,Kuijjer:2018}.  In this context, the aim is to identify clusters of patients on the basis of a diverse range of molecular variables, measurements of which are obtained using high-throughput 'omics technologies.  One feature of these datasets is that they are typically high-dimensional, which frequently necessitates the use of variable screening or selection strategies \citep[e.g.][]{Witten:2010,Fop:2018,Crook:2018}, or other dimension reduction techniques \citep[e.g.][]{Yeung:2001,McLachlan:2002,Taschler:2019}.  However, a commonly overlooked challenge is that 'omics datasets often define {\em multiple} clustering structures in the patient population, as a consequence of different subsets of variables (e.g. those corresponding to functional groups of genes or proteins) being implicated in a variety of different biomolecular processes.  Thus, depending on the subset of variables we consider, we can identify different patient clusters.  

A number of papers have proposed methods for identifying multiple clustering structures \citep[e.g.][]{Cui2007,Niu2010,Guan2010,Li2011,Niu2014}.  In the literature, a set of variables that define the same clustering structure has been termed a {\em view}, while the task of identifying views and their associated clustering structures has been referred to as either {\em multi-view clustering} \citep{Cui2007} or {\em cross-clustering} \citep{Li2011}.  One potential challenge faced by these approaches is how to decide which of the identified clustering structures is the most useful or relevant for a given task.  This is particularly important in stratified medicine and disease subtyping applications, where we seek clusters (strata) that define groups of patients who have, for example, similar prognoses or respond similarly to treatment \citep[e.g.][]{Perou2000,Sorlie2001}.  In practice, to determine if a given clustering structure is ``relevant'', it is common to make use of a left out outcome variable $y$ (e.g. survival data) and to assess whether or not different clusters are associated with different distributions of $y$ \citep[e.g.][]{Curtis:2012}.  An alternative approach is to adopt a semi-supervised (outcome guided) approach that makes use of $y$ when performing the clustering analysis.     

{\em Bayesian profile regression} is one such semi-supervised mixture modelling approach that makes use of an outcome/response in order to guide inference toward relevant clustering structures \citep{Molitor2010}.  Informally, a clustering is said to be relevant (for a given response) if individuals allocated to the same cluster tend to have similar values for the response; or, more generally, if the responses of individuals in the same cluster can be accurately described by a common model.  More precisely, a clustering is defined to be relevant (for a given response) if the value taken by an individual's response is not independent of their cluster allocation. It is clear from this definition that the relevance of a clustering can only be specified relative to a given response -- and, in particular, that for different responses, different clusterings might be relevant.     

For example, we could use Bayesian profile regression to retrospectively cluster patients on the basis of genetic or genomics data, using their survival times as a response to guide the clustering toward prognostically relevant disease subtypes.  If we were to use a different response (e.g. height), we might end up with a completely different clustering structure.  Neither one of these clustering structures would necessarily be ``wrong" -- they are just relevant with respect to different responses.  Crucially, if we were to adopt an {\em unsupervised} clustering approach (which is commonly the default analysis choice), there is no guarantee that this would identify a clustering structure that was relevant for either response.  

As we demonstrate in Section \ref{simstudy}, a limitation of the Bayesian profile regression model is that, with increasing data dimension, the influence exerted by the (typically low dimensional) response on the inference of the clustering structure grows weaker.  Here we propose a semi-supervised multi-view Bayesian clustering model that simultaneously addresses both this limitation, as well as the challenge faced by existing multi-view approaches of picking out the (most) relevant clustering structure. 


\section{Profile regression}\label{profile_regression} 
We suppose that we have data comprising observations on a vector of clustering variables, ${\bf x}$, and responses,~${\bf y}$.  We denote the concatenated vector of clustering variables and response by ${\bf v} = [{\bf x}; {\bf y}]$, and  
 model the data using a mixture model with $K$ components (where $K$ could be finite or infinite), as follows:
\begin{align}
p({\bf v}|\boldsymbol{\rho}, \boldsymbol{\pi}) & = \sum_{k=1}^K \pi_k f_{\bf v}({\bf v}|\boldsymbol{\rho}_k)\\
&= \sum_{k=1}^K \pi_k f_{\bf y}({\bf y}|\boldsymbol{\theta}_k, {\bf x})f_{\bf x}({\bf x}|\boldsymbol{\phi}_k),\label{factored}
\end{align}
where $\pi_k$ is the mixture weight associated with the $k$-th component, $\boldsymbol{\rho}_k$ denotes the parameters associated with the $k$-th component, and we write $\boldsymbol{\rho}$ and $\boldsymbol{\pi}$ to denote $\{\boldsymbol{\rho}_k \}_{k = 1}^K$ and $\{\pi_k \}_{k = 1}^K$ respectively.  We assume that the joint density $f_{\bf v}$ can be factorised into $f_{\bf x}$ and $f_{ {\bf y}}$ as shown in Equation~\eqref{factored}, with $\boldsymbol{\phi}_k$ denoting the parameters of the model for~${\bf x}$, and $\boldsymbol{\theta}_k$ denoting the parameters of the model for~${\bf y}$.  We will write $\boldsymbol{\Phi}$ and $\boldsymbol{\Theta}$ to denote $\{\boldsymbol{\phi}_k \}_{k = 1}^K$ and $\{\boldsymbol{\theta}_k \}_{k = 1}^K$ respectively.  In the profile regression model, it is further assumed that $f_{\bf y}({\bf y}|\boldsymbol{\theta}_k, {\bf x}) = f_{\bf y}({\bf y}|\boldsymbol{\theta}_k)$; i.e. that ${\bf y}$ is conditionally independent of ${\bf x}$ given $\boldsymbol{\theta}_k$ \citep{Molitor2010}.  A related model, in which this conditional independence assumption is not made, is given in \citet{Shahbaba2009}.  

In \citet{Molitor2010}, the authors allow the model for ${\bf y}$ to include a dependence upon additional adjustment covariates, ${\bf w}$, that may be predictive of the response but that we do not wish to contribute to the clustering (e.g. confounders that we wish to control for, such as age or sex), together with associated ``global" (i.e. not component-specific) parameters $\boldsymbol{ \beta}$ (see Appendix for details).  The general profile regression model is then:
\begin{align}
p({\bf x}, {\bf y}|\boldsymbol{\Phi}, \boldsymbol{\Theta}, \boldsymbol{\pi}, \boldsymbol{\beta}, {\bf w}) = \sum_{k=1}^K \pi_k f_{\bf y}({\bf y}|\boldsymbol{\theta}_k, {\bf w}, \boldsymbol{ \beta})f_{\bf x}({\bf x}|\boldsymbol{\phi}_k).\label{generalProfile}
\end{align}

The original formulation of the profile regression model, which we also adopt here, is specifically in terms of infinite mixture models using Dirichlet process priors \citep{Molitor2010}; however, we note that the model is equally applicable in the case of finite $K$.

\subsection{Dirichlet process formulation}
In \citet{Molitor2010}, a finite approximation to the Bayesian nonparametric case was considered, using a truncated stick breaking construction of the Dirichlet process \citep{Ishwaran2001} to define the prior on the $\pi_k$'s in Equation \eqref{generalProfile}.  Subsequent Bayesian profile regression papers \citep{Hastie2014} and implementations \citep{Liverani2015} also considered stick breaking constructions, with inference performed via slice sampling \citep{Walker2007,Kalli2011}.  Following the derivations of \citet{Neal2000} and \citet{Rasmussen2000} in the unsupervised case, here we instead consider the Dirichlet process mixture model as a limiting case of a (finite) $K$ component mixture model, in which a symmetric Dirichlet prior with parameter $\alpha/K$ is placed on the mixture weights, $\pi_k$ \citep[see also][]{Ishwaran2002}.  This is closely related to the P\'{o}lya urn \citep{Blackwell1973} and Chinese restaurant process \citep{Aldous1985} constructions for the Dirichlet process, and permits inference via a collapsed Gibbs sampler in which the mixture weights are marginalised.  Details are provided in the Appendix, with sampling performed as in \citet{Neal2000} -- although we note that alternative approaches for performing inference in Bayesian mixture models could also be employed in this context \citep[e.g.][]{Richardson1997,Green:2001,Jain:2004,Jain:2007,Walker2007,Kalli2011,Miller:2017} 

To provide a very brief overview, let $D = \{({\bf x}_i, {\bf y}_i, {\bf w}_i) \}_{i=1}^n$ denote a dataset comprising $({\bf x}, {\bf y}, {\bf w})$ triples for $n$ individuals, such that $({\bf x}_i, {\bf y}_i, {\bf w}_i)$ corresponds to the $i$-th individual.   As is common for mixture models, we introduce latent component allocation variables, $z_i$, where $z_i = k$ if the $i$-th individual is associated with the $k$-th component, and $p(z_i = k| \boldsymbol{\pi}) = \pi_k$.  We define ${\bf z} = \{ z_1, \ldots, z_n\}$ to be the multiset of all $n$ component allocations.  The component-conditional likelihood associated with the $i$-th individual is then:
\begin{align}
p({\bf x}_i,\mathbf{y}_i |z_i = k, \boldsymbol{\phi}, \boldsymbol{\theta}, \boldsymbol{\beta}, {\bf w}_i) = f_{\bf y}({\bf y}_i|\boldsymbol{\theta}_k, {\bf w}_i, \boldsymbol{ \beta})f_{\bf x}({\bf x}_i | \boldsymbol{\phi}_{k}).
\end{align}

Independent priors, $p(\boldsymbol{\theta}), p(\boldsymbol{\phi})$ and $p(\boldsymbol{\beta})$, are taken for the component-specific and global parameters.  The collapsed Gibbs sampler then iterates between: 
\begin{itemize}
\item Step (i): updating the component allocations, $z_i$, given the data $D$, the most recently sampled parameters $\boldsymbol{\Theta}, \boldsymbol{\Phi},\boldsymbol{\beta}$, and the most recently sampled values for the other allocation variables, ${\bf z}_{-i}= {\bf z}\backslash \{ z_i\}$; and 
\item Step (ii): updating the parameters $\boldsymbol{\Theta}, \boldsymbol{\Phi},\boldsymbol{\beta}$, given the data $D$ and the most recently sampled component allocations~${\bf z}$.  
\end{itemize}

In the finite $K$ case, if we take a symmetric Dirichlet prior with parameter $\alpha/K$ for the mixture weights, $\pi_1, \ldots, \pi_K \sim \mbox{Dir}(\alpha/K)$, then the conditional posterior probability of allocating individual~$i$ to the $k$-th component -- required for Step (i) above -- is given by:
\begin{align}
p(z_i=k |D,&{\bf z}_{-i}, \boldsymbol{\Phi}, \boldsymbol{\Theta}, \boldsymbol{\beta}, {\bf w}_i, \alpha) = b\frac{n_{-i,k}}{n - 1 + \alpha} f_{\bf y}({\bf y}_i|\boldsymbol{\theta}_k, {\bf w}_i, \boldsymbol{ \beta})f_{\bf x}({\bf x}_i | \boldsymbol{\phi}_{k}), \label{mycond1}
\end{align}
where $b$ is a normalising constant that ensures that $\sum_{k = 1}^K p(z_i=k |D,{\bf z}_{-i}, \ldots) = 1$, and $n_{-i,k}$ is the number of individuals currently allocated to component $k$, excluding the $i$-th individual.  That is, if we let $\mathbbm{1}_a$ denote the indicator function (which is equal to 1 if $a$ is true and zero otherwise), then $n_{-i,k} = \sum_{z_j \in {\bf z}_{-i}}\mathbbm{1}_{z_j = k}$.

The Dirichlet process (DP) mixture model may be derived by considering the limit $K \rightarrow \infty$, in which case the conditional posterior probability of allocating individual $i$ to an existing component (i.e. one to which other individuals are currently allocated) is given by:
%
%
%
\begin{align}
p(z_i = k, \mbox{ where } k \in {\bf z}_{-i} |D,&{\bf z}_{-i}, \boldsymbol{\Phi}, \boldsymbol{\Theta}, \boldsymbol{\beta}, {\bf w}_i, \alpha)= b\frac{n_{-i,k}}{n - 1 + \alpha} f_{\bf y}({\bf y}_i|\boldsymbol{\theta}_k, {\bf w}_i, \boldsymbol{ \beta})f_{\bf x}({\bf x}_i | \boldsymbol{\phi}_{k}), \label{mycond1}
\end{align}
where $n_{-i,k}$ is as before and $b$ is again a normalising constant (see Equation \eqref{normConst} below).  For the DP, we also require the conditional posterior probability of allocating individual $i$ to a new component, which is:
\begin{align}
p(z_i \not \in {\bf z}_{-i} |D,&{\bf z}_{-i}, \boldsymbol{\Phi}, \boldsymbol{\Theta}, \boldsymbol{\beta}, {\bf w}_i, \alpha)= b\frac{\alpha}{n - 1 + \alpha} \int_{\boldsymbol{\phi}, \boldsymbol{\theta}}f_{\bf y}({\bf y}_i|\boldsymbol{\theta}, {\bf w}_i, \boldsymbol{ \beta})f_{\bf x}({\bf x}_i | \boldsymbol{\phi}) p(\boldsymbol{\phi})p(\boldsymbol{\theta}) d\boldsymbol{\phi} d\boldsymbol{\theta}.\label{mycond2}
\end{align}

The normalising constant, $b$, in Equations \eqref{mycond1} and \eqref{mycond2} is chosen to ensure that: 
\begin{equation}
p(z_i \not \in {\bf z}_{-i} |D,{\bf z}_{-i}, \ldots) + \sum_{k \in {\bf z}_{-i}} p(z_i = k, \mbox{ where } k \in {\bf z}_{-i} |D,{\bf z}_{-i}, \ldots) = 1. \label{normConst}
\end{equation}

Given values for the component allocation variables, ${\bf z}$, it is relatively straightforward to perform Step (ii); i.e. to sample the parameters $\boldsymbol{\Theta}, \boldsymbol{\Phi},\boldsymbol{\beta}$. See Appendix for full details, where we also describe how to sample the DP hyperparameter, $\alpha$, according to the method described in \citet{Escobar1995}.



\section{Variable selection in profile regression}\label{variable_selection}
The multi-view model that we propose in Section \ref{multi-view} may be regarded as an extended form of variable selection for clustering.  We therefore start by presenting a version of the profile regression model that permits variable selection in the same manner as in the models proposed by \citet{Law2003,Law2004,Tadesse2005,Papathomas2012}.  


We again consider the component-conditional likelihood associated with the $i$-th individual, making the additional assumption of conditionally independent clustering variables:
\begin{align}
f(\mathbf{x}_i,\mathbf{y}_i | z_i = k, \boldsymbol{\phi}, \boldsymbol{\theta}, \boldsymbol{\beta}, {\bf w}_i) &= f_\mathbf{y}({\bf y}_i|\boldsymbol{\theta}_k,\boldsymbol{\beta},\mathbf{w}_i)f_\mathbf{x}(\mathbf{x}_i | \boldsymbol{\phi}_k)\notag\\
&= f_{\bf y}({\bf y}_i|\boldsymbol{\theta}_k,\boldsymbol{\beta},\mathbf{w}_i)\prod_{j=1}^P f_x(x_{ij} | \boldsymbol{\phi}_k)\label{profRegr},\vspace{-0.5cm}
\end{align}
where $P$ is the dimension of ${\bf x}$, so ${\bf x}_i = [x_{i1}, \ldots, x_{iP}]^\top$.  In the following, we assume throughout a common parametric form, $f_x$, for all clustering variables; however, we could straightforwardly extend to allow different parametric forms for different clustering variables in order to allow, for example, modelling of mixed (continuous and categorical) data.

In order to perform variable selection, we follow the same approach as in \citet{Law2003,Law2004} and \citet{Tadesse2005}, and introduce latent binary indicator variables $\gamma_j$, such that $\gamma_j = 1$ if the $j$-th clustering variable contributes to the clustering structure, and 0 otherwise.  We model the clustering variables for which $\gamma_j = 0$ as having density $q_{x}(x_j | \boldsymbol{\phi}_{0j})$, where $\boldsymbol{\phi}_{0j}$ is a ``global" (i.e. not component-specific) parameter and $q_x$ need not be of the same parametric form as $f_x$.  We denote the collection of all $\boldsymbol{\phi}_{0j}$ parameters by $\boldsymbol{\phi}_{0\cdot} = \{ \boldsymbol{\phi}_{0j}\}_{j = 1}^P$. The allocation-conditional likelihood is then:
\begin{align}
f(\mathbf{x}_i,y_i | z_i = k, \boldsymbol{\phi}, \boldsymbol{\phi}_{0\cdot}, \boldsymbol{\theta}, {\bf w}_i, \boldsymbol{\beta},\boldsymbol{\gamma}) &=f_\mathbf{y}({\bf y}_i|\boldsymbol{\theta}_k,\boldsymbol{\beta},\mathbf{w}_i) f_\mathbf{x}(\mathbf{x}_i | \boldsymbol{\phi}_k)\notag\\
&= f_{\bf y}({\bf y}_i|\boldsymbol{\theta}_k,\boldsymbol{\beta},\mathbf{w}_i)\prod_{j=1}^P q_x(x_{ij} | \boldsymbol{\phi}_{0j})^{\mathbbm{1}_{\gamma_j = 0}} f_x(x_{ij} | \boldsymbol{\phi}_k)^{\mathbbm{1}_{\gamma_j = 1}} .\label{vblSelect}\vspace{-0.5cm}
\end{align}

The introduction of the latent indicator variables in Equation \eqref{vblSelect} results in the grouping of clustering variables into two disjoint sets.  Adopting the terminology of \citet{Cui2007}, we will refer to these sets as {\em views}.  We define View~1 to be the set of clustering variables (for which $\gamma_j = 1$) that define a clustering structure, and View~0 to be the set (for which $\gamma_j = 0$) that do not define a clustering structure.  We will refer to this latter set of clustering variables as the {\em null view}. The variable selection model may therefore be considered to be one which not only groups individuals together (into clusters), but also groups clustering variables together (into 2~views).  Inference of the view allocation variables, $\gamma_j$, can be performed via Gibbs sampling, and amounts to sampling according to the posterior probabilities associated with each of the models (i.e. the ``no clustering structure" model, $q_x$, and the ``clustering structure" model, $f_x$) given the observed data for the $j$-th clustering variable, $\{ x_{ij}\}_{i=1}^n$.  See Appendix for details.


\section{Multi-view Bayesian profile regression}\label{multi-view}
A natural extension to the variable selection model is to allow there to be $L$ views (with $L \in \mathbb{Z}^+$ now allowed to be more than~2) by allowing the indicator variables, $\gamma_j$, to be categorical with $L$ categories, $0, 1, 2, \ldots, L-1$.  It is then clear that the $\gamma_j$ variables are {\em view allocation variables}, which serve an analogous role to the component allocation variables, $z_i$, but which act to group together the clustering variables rather than the individuals.

As in the variable selection model, the introduction of multiple views necessitates the introduction of view-specific models.  In the variable selection case, the clustering variables in View 1 contribute to a mixture model (which defines clusters among the individuals), while the clustering variables in the null view contribute to a single density (corresponding to all individuals being in a single cluster).  In general, different choices are possible for the view-specific models.  Here we consider an $L$-view case in which View~0 is a null view, while Views $1, \ldots, L-1$ are each associated with a mixture model, such that each of these views defines a different clustering structure among the individuals.  We associate the response with View~1 only, and refer to this view as the {\em relevant view}.  Thus, the View~1 model is a (semi-supervised) Bayesian profile regression model, while the models for Views $2, \ldots, L - 1$ are all unsupervised mixture models.  

\subsection{Allocations-conditional likelihood}       
For each of the non-null views, we introduce component allocation variables $z_{i\ell}$, such that $z_{i\ell}$ denotes the component that is responsible for the $i$-th individual in the $\ell$-th view's mixture model.  We denote the number of mixture components in the $\ell$-th view's mixture model by $K_\ell$, where -- as previously -- $K_\ell$ may be finite or infinite.  We moreover denote the parameters associated with the $k$-th component in the $\ell$-th view by $\boldsymbol{\phi}_{\ell k}$ and define $\boldsymbol{\phi}_{\ell\cdot} = \{ \boldsymbol{\phi}_{k\ell}\}_{k = 1}^{K_\ell}$ to be the full complement of component-specific parameters associated with the $\ell$-th view.  The $\boldsymbol{\theta}$ parameters of the model for~${\bf y}$ are associated with view $\ell = 1$ only.

The likelihood associated with the $i$-th individual, conditioned on these component allocations, is then:
\begin{align}
f(\mathbf{x}_i,y_i | \{z_{i\ell} = k_\ell, \boldsymbol{\phi}_{\ell\cdot}\}_{\ell=1}^{L-1}, \boldsymbol{\phi}_{0\cdot}, \boldsymbol{\theta}, &{\bf w}_i, \boldsymbol{\beta},\boldsymbol{\gamma}) 
=\notag\\ 
&f_{\bf y}({\bf y}_i|\boldsymbol{\theta}_{k_1},\boldsymbol{\beta},\mathbf{w}_i)\prod_{j=1}^P q_x(x_{ij} | \boldsymbol{\phi}_{0j})^{\mathbbm{1}_{\gamma_j = 0}} \prod_{\ell=1}^{L-1} f_x(x_{ij} | \boldsymbol{\phi}_{\ell k_\ell})^{\mathbbm{1}_{\gamma_j = \ell}} .\label{generalModel}\vspace{-0.5cm}
\end{align}     

\begin{figure}[h!]\centering
{\includegraphics[width=0.7\linewidth,trim = 0mm 0mm 0mm 0mm, clip]{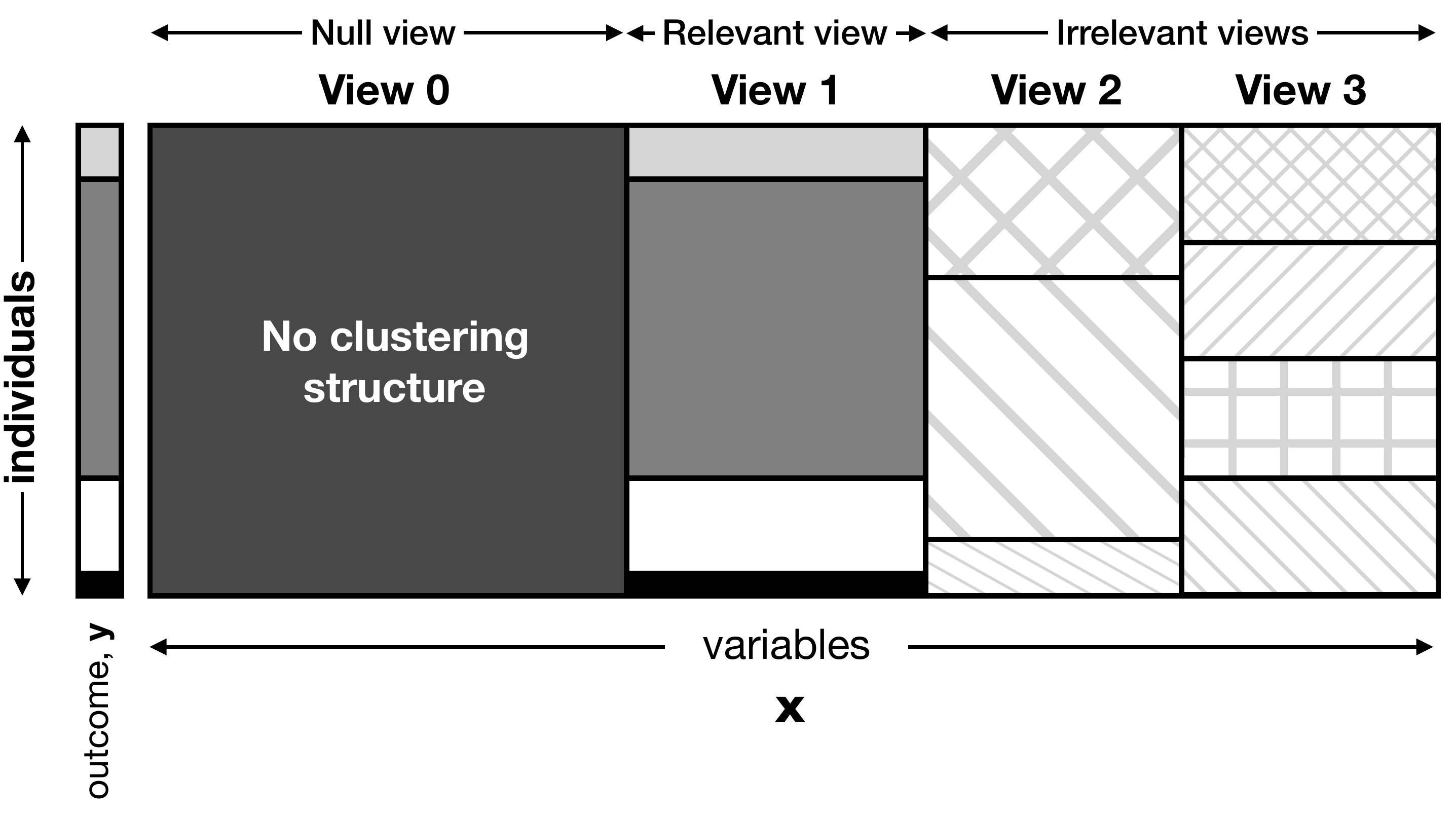}}\caption{Schematic illustration of semi-supervised multi-view model.  Conditioned on the allocation of clustering variables to views, each view is modelled independently, with View~0 (the null view) modelled as possessing no clustering structure, View~1 modelled with a profile regression (semi-supervised clustering) model that links the clustering structure defined by the clustering variables to a response, $\bf{y}$, and the remaining views each modelled by their own unsupervised mixture model.  The level of the response, ${\bf y}$, is indicated by a column shown to the left of the data matrix in the figure above. }\label{multi-viewFigure}
\end{figure}

Defining $\Gamma_\ell = \{j : \gamma_j = \ell \}$ to be the index set of clustering variables allocated to the $\ell$-th view, the above may alternatively be written as follows:
\begin{align}
f(\mathbf{x}_i,y_i | &\{z_{i\ell} = k_\ell, \boldsymbol{\phi}_{\ell\cdot}\}_{\ell=1}^{L-1}, \boldsymbol{\phi}_{0\cdot}, \boldsymbol{\theta}, {\bf w}_i, \boldsymbol{\beta},\boldsymbol{\gamma}) 
=\notag\\ 
&\blackub{\left(\prod_{j\in \Gamma_0} q_x(x_{ij} | \boldsymbol{\phi}_{0j})\right)}_{\mathclap{\text{View 0}}}  \blackub{\left(f_{\bf y}({\bf y}_i|\boldsymbol{\theta}_{k_1},\boldsymbol{\beta},\mathbf{w}_i) \prod_{j\in \Gamma_1} f_x(x_{ij} | \boldsymbol{\phi}_{1 k_1})   \right)}_{\mathclap{\text{View 1}}}   \prod_{\ell=2}^{L-1} \blackub{\left(\prod_{j\in \Gamma_\ell} f_x(x_{ij} | \boldsymbol{\phi}_{\ell k_\ell})\right)}_{\mathclap{\text{View $\ell$}}}, \label{GammaNotation}\vspace{-0.5cm}
\end{align}
where each of the bracketed terms corresponds to a different view, as shown.  This expression makes clear that (conditioned on the allocation of clustering variables to views) each of the views is modelled independently, as also illustrated in Figure~\ref{multi-viewFigure}, with View~0 being modelled as a single cluster, View~1 being modelled with a profile regression model (compare to Equation \eqref{profRegr}), and the remaining views each modelled by their own (unsupervised) mixture model.  

\subsection{Inference}
Conditioned on the view allocations, the models describing each of the views are independent, and hence we can perform inference for the component allocations and parameters within each view using existing approaches either for Bayesian profile regression models \citep{Molitor2010} in the case of View 1, or for unsupervised Bayesian mixture models \citep[e.g.][]{Richardson1997,Neal2000} for Views 2, \ldots, $L-1$.  We therefore adopt a Gibbs sampling approach, in which we iterate between sampling the view allocation variables and performing inference for the models in each view.  We provide details of the update for the view allocation indicators below.  Updates for the within-view parameters and latent variables are performed as in \citet{Neal2000}; see  Appendix for further details.

\subsubsection{Updating the view allocation indicators}  
Define $\nu_{\ell}$ to be the prior probability that a clustering variable is allocated to the $\ell$-th view.  Given the component allocations $z_{\cdot\ell} = \{z_{i\ell} \}_{i = 1}^n$ for the $\ell$-th view (for $\ell = 1, \ldots, L-1$), the posterior probability that the $j$-th clustering variable is allocated to the $\ell$-th view is then:
\begin{align}
p(\gamma_j = \ell | z_{\cdot\ell}, \boldsymbol{\phi}_{\ell \cdot} ) = \frac{1}{Z} \nu_{\ell} \prod_{i = 1}^n f_x(x_{ij} | \boldsymbol{\phi}_{\ell z_{i\ell}}) \mbox{ for $\ell = 1, \ldots, L-1$},
\end{align}
where $Z$ is a normalising constant that ensures that the posterior view allocation probabilities sum to 1, and, as before, $\boldsymbol{\phi}_{\ell z_{i\ell}}$ denotes the parameter associated with the $z_{i\ell}$-th component in the $\ell$-th view, and $\boldsymbol{\phi}_{\ell \cdot}$ is the full complement of component-specific parameters associated with the $\ell$-th view.

For the null view, we have:
\begin{align}
p(\gamma_j = 0|  \boldsymbol{\phi}_{0 j} ) = \frac{1}{Z}  \nu_{0} \prod_{i = 1}^n q_x(x_{ij} | \boldsymbol{\phi}_{0j}),
\end{align}
and it is now clear that $$Z = \nu_{0} \prod_{i = 1}^n q_x(x_{ij} | \boldsymbol{\phi}_{0j}) + \sum_{\ell = 1}^{L-1}\left( \nu_{\ell} \prod_{i = 1}^n f_x(x_{ij} | \boldsymbol{\phi}_{\ell z_{i\ell}})\right).$$

In the case where conjugate priors are taken for the component-specific parameters, $\boldsymbol{\phi}_{\ell k}$, and/or the parameters of the null view model, $\boldsymbol{\phi}_{0.}$, these parameters may be integrated out and the likelihood functions $f_x$ and $q_x$ may be replaced with marginal likelihoods; see Appendix.

\subsection{Choice of $L$, the number of views}
It may be noted that inference for the view allocations is somewhat analogous to inference for the mixture component allocations within each (non-null) view.  In the latter case, individuals are allocated with a higher probability to components in which individuals with similar clustering variable profiles are allocated; whereas in the former case, clustering variables are allocated with a higher probability to views in which clustering variables defining a similar clustering structure are allocated.    

Similarly, the choice of the number of views in a multi-view model, $L$, is analogous to the choice of the number of components, $K$, in a conventional mixture model.  Moreover, in much the same way that the prior component allocation probabilities in a mixture model may be treated as parameters, $\pi_k$, to be inferred (e.g. taking a Dirichlet or Dirichlet process prior), it is also possible to treat the prior view allocation probabilities in the multi-view model, $\nu_\ell$, as parameters.  By adopting a Dirichlet (or Dirichlet process) prior, and taking $L$ to be large (or infinite), the number of views may, in principle, be inferred automatically.  However, this comes with an associated computational expense, since each additional view brings with it a mixture model whose parameters and latent component allocations must be inferred.  Approximate inference procedures for these models, such as variational Bayes \citep[previously considered in the context of multi-view clustering by][]{Guan2010} will be an important direction for future research, but in the present work we focus upon small values of $L \ge 2$, for which inference via Markov chain Monte Carlo (MCMC) is feasible.   

\subsection{Initialisation}
We have found that a good initialisation strategy is to start with all variables in the relevant view, so that irrelevant and null variables are ``selected out" at subsequent iterations.  We adopt this initialisation strategy in all examples   

\section{Examples}
Although until now we have deliberately kept the exposition general, in the examples that follow we restrict our attention to Dirichlet process mixture models for the non-null views.  In Section \ref{simstudy} we present simulation examples that allow us to illustrate how clustering approaches that do not model multiple clustering structures can fail, even if they exploit response information and employ variable selection.  In Section \ref{brcaexample} we consider an integrative clustering example in the context of breast cancer subtype characterisation, in which we fit a multi-view model with 3 views (including one null view).  

\subsection{Simulation study to illustrate the limitations of methods that ignore multiple views}\label{simstudy}
\begin{figure}[h]
    \centering
    \begin{subfigure}[b]{.48\textwidth}
\PPPlotFrame{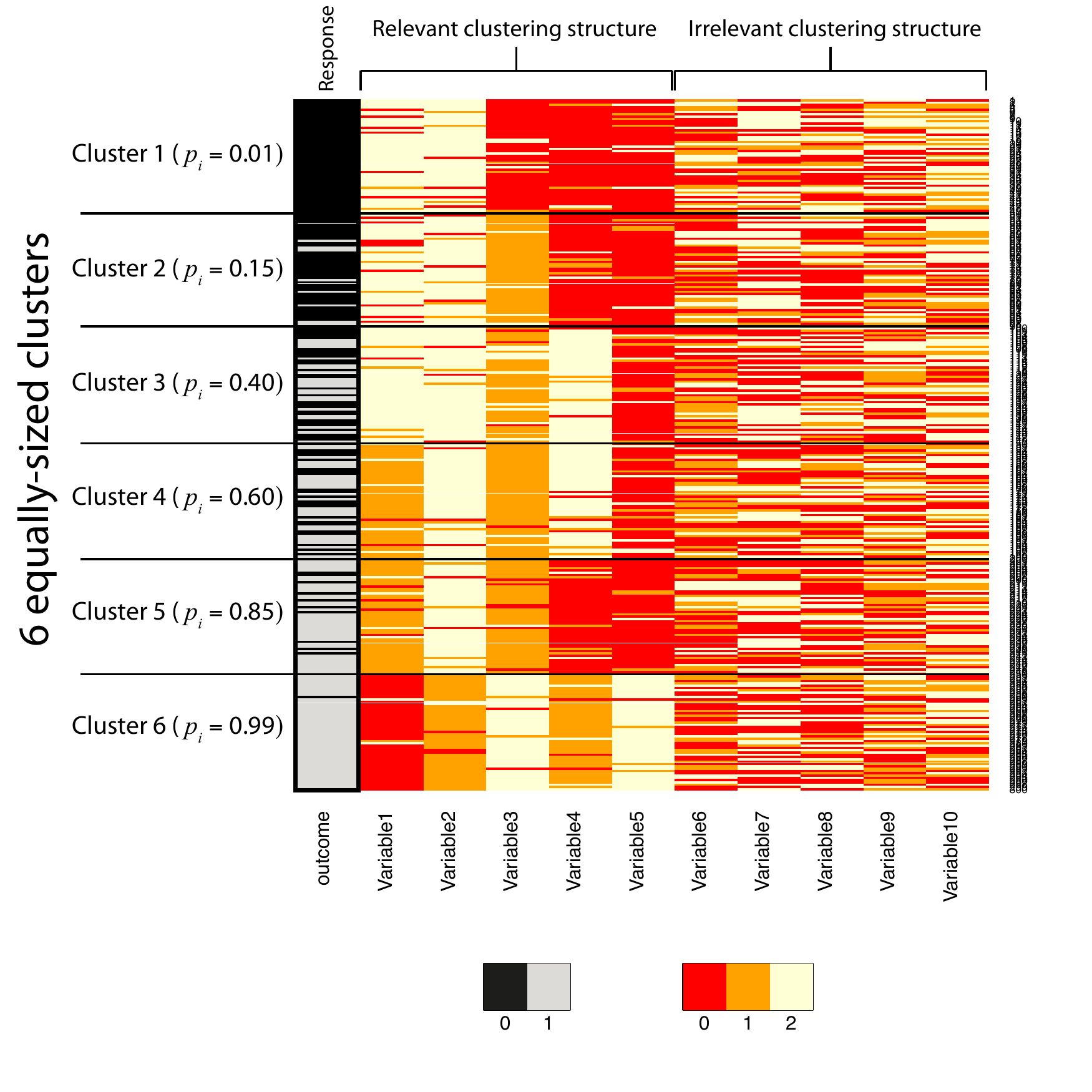}
        \caption{Rows ordered to emphasise the {\em relevant} clustering structure.}
        \label{figSimExA}
    \end{subfigure}
    ~ 
    \begin{subfigure}[b]{.48\textwidth}
\PPPlotFrame{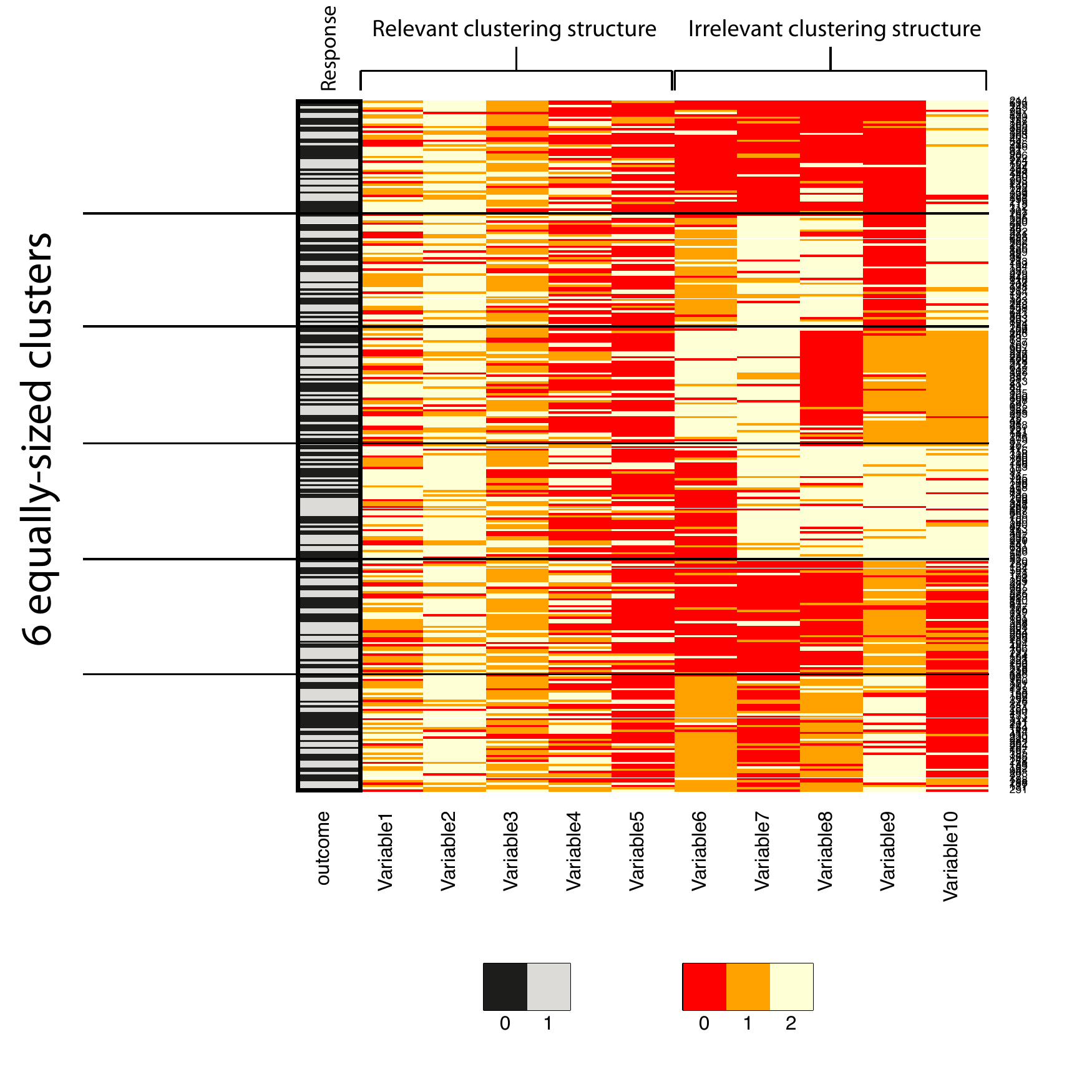}
        \caption{Rows ordered to emphasise the {\em irrelevant} clustering structure.}
        \label{figSimExB}
    \end{subfigure}
    \caption{The same simulated dataset with 2~different row orderings.  The data within each view are simulated according to Equation~\eqref{simModel}, with $w = 0.8$. (a) The rows are ordered to show the clusters defined by the clustering variables in the relevant view (variables 1 to 5).  The probability that an individual's response is 0/1 depends upon their membership of these clusters.  (b)  As in (a), but ordering the rows to highlight the clusters defined by the clustering variables in the irrelevant view (variables 6 to 10).  As described in the main text, there is no link between these (irrelevant) clusters and the response. }\label{figSimEx}
\end{figure}

We construct a simulation example to demonstrate the limitations of existing semi-supervised clustering approaches that fail to account for multiple clustering structures.  We consider simulated datasets with $n = 300$ individuals, $p = 10$ categorical clustering variables (each of which has 3 categories), and a univariate binary response,~$y$.  The clustering variables define 2 views, with the first $q$ clustering variables ($q \in \{3, 5, 7, 9\} $) defining a {\em relevant} clustering structure  (which relates to the response) and the remaining $p-q$ defining an {\em irrelevant} clustering structure (unrelated to the response).  The clustering variables in the relevant view (View 1) define 6 equally-sized clusters, which are related to the response according to $P(y_i = 1 | z_{i1} = k) = \theta_k$, where $\theta_k \in \{0.01, 0.15, 0.40, 0.60, 0.85, 0.90\}$.  The clustering variables in the irrelevant view (View 2) also define 6 clusters, but there is no link between these clusters and the response; i.e. $P(y_i = 1 | z_{i2} = k) = P(y_i = 1 ) = 0.485$, regardless of the component allocation in View 2.  An illustration of a dataset for $q = 5$ is provided in Figure~\ref{figSimEx}, with Figure~\ref{figSimExA} showing the clustering structure defined by the clustering variables in View 1, and Figure~\ref{figSimExB} showing the clustering structure defined by the clustering variables in View 2 (which is irrelevant for the response shown).  

Within each view, the categorical data are simulated according to a mixture distribution.  With probability $w \in [0,1]$ we simulate $x_{ij}$ according to:
\begin{align}
x_{ij} | z_i = k, \gamma_j = \ell &\sim \textrm{Categorical}(\boldsymbol{\phi}_{\ell kj}),\notag\\
\mbox{ where }\boldsymbol{\phi}_{\ell kj} = [{\phi}_{\ell kj1}, {\phi}_{\ell kj2}, {\phi}_{\ell kj3}] &\sim \textrm{Dirichlet}(0.01, 0.01, 0.01),\label{simModel}
\end{align} 
and with probability $1 - w$ we simulate $x_{ij}$ according to a discrete uniform distribution on the three categories.  The parameter $w$ thereby allows us to control cluster separability.  The clusters are easily separable when $w$ is close to 1,  but the clusters are increasingly noisy and hard to separate as $w$ approaches 0. 

\subsubsection{Results}\label{simresults}
 We applied an existing semi-supervised clustering approach to each dataset, as implemented in the PReMiuM R package for Bayesian profile regression \citep{Liverani2015}.  For each dataset, we ran PReMiuM both with and without variable selection.  To run with variable selection, we specified the ``varSelectType" option in PReMiuM to be ``continuous", which performs variable selection with latent selection weights as described in \citet{Papathomas2012}.  To obtain a ``gold standard" that reflects the performance that could be achieved if only the relevant clustering variables were selected, we additionally applied PReMiuM to each dataset after having removed all of the irrelevant clustering variables.  In practical examples, identifying the relevant clustering variables a priori in this way will not be possible; however, here we suppose we have access to an oracle that can provide this information.  In all cases, we used PReMiuM to perform 10,000 Gibbs sampling iterations, discarding the first 1,000 as burn in and thinning to retain every 5-th draw.

To summarise the MCMC output, we calculated the adjusted Rand index (ARI) between the true clustering structure in the relevant view and each retained clustering structure (partition) sampled from the posterior.  Thus, for each distinct PReMiuM run, we obtained a distribution of ARI values, which assesses how well the inferred clustering structure matches the clustering structure in the relevant view.  The ARI can take a value of at most 1 (indicating a perfect match), while a value of 0 indicates that the match is no better than we would expect by random chance.  

\begin{figure}[!h]
    \centering
\includegraphics[width = 0.8\linewidth]{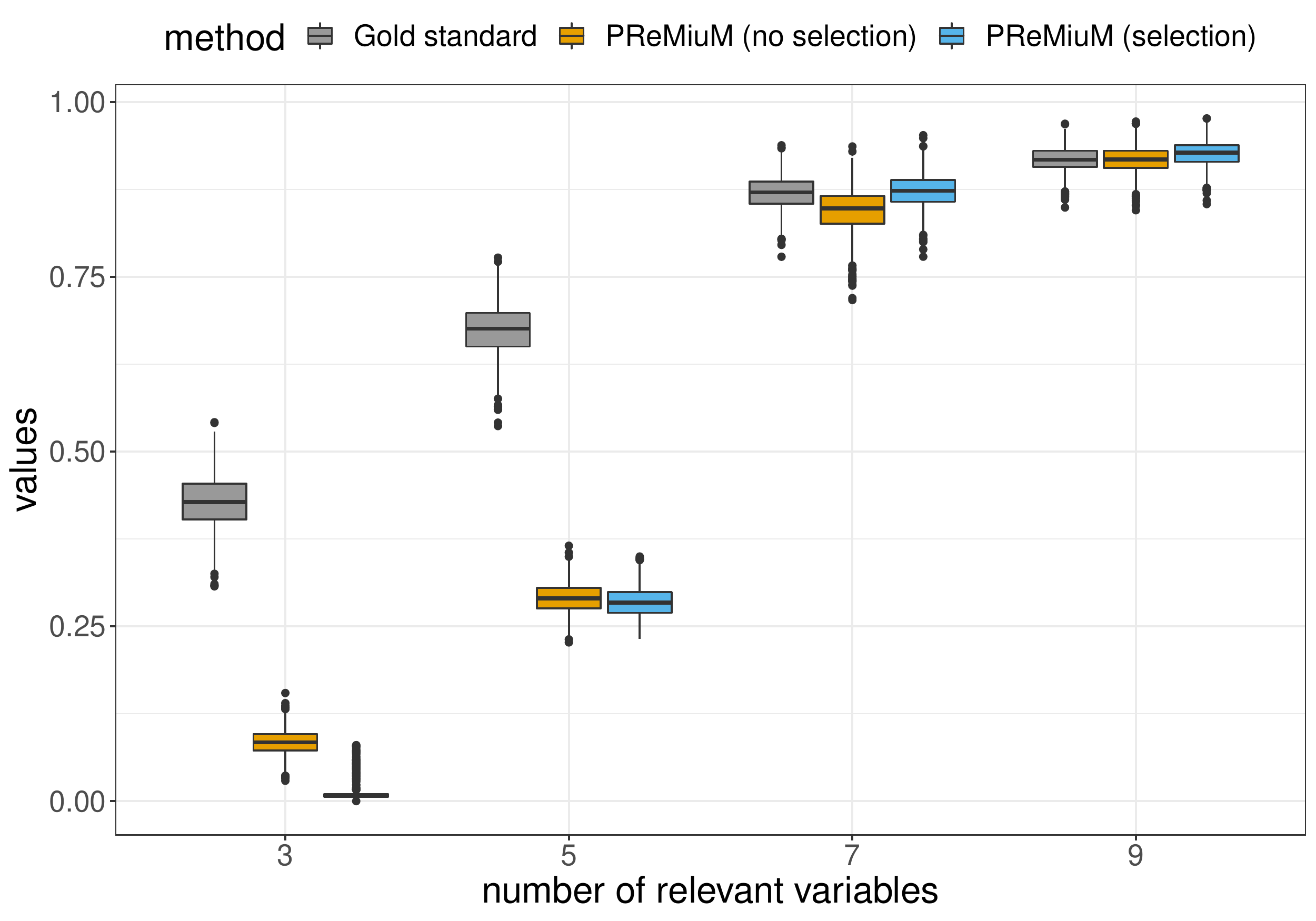}
        \caption{Posterior distributions of Adjusted Rand index (ARI) values obtained when applying PReMiuM to datasets comprising a mix of relevant and irrelevant clustering variables.  PReMiuM is applied either with (blue) or without (gold) variable selection to datasets comprising both relevant and irrelevant variables, or to the same datasets with the irrelevant variables removed (grey; ``gold standard" performance).  
        Solid lines: applying PReMiuM without variable selection to each dataset.  Dashed lines: applying PReMiuM with variable selection to each dataset.  Dotted lines: applying PReMiuM to each dataset after having artificially removed the irrelevant clustering variables.}
        \label{fig03}
\end{figure}

Figure~\ref{fig03} illustrates typical output for 4 datasets, each having a different value for $q$ (the number of relevant clustering variables).  Corresponding figures for other simulated datasets are provided in Supplementary Results, and are qualitatively similar.  As we might expect, as $q$ diminishes, so too does our ability to infer the correct clustering structure.  Perhaps less intuitively, we also see that -- as a consequence of failing to model the multiple views --  running PReMiuM with variable selection does not necessarily help to improve inference of the relevant clustering structure, and can actually be damaging.  Figure~\ref{fig03} shows that variable selection can be useful when the number of relevant clustering variables is large relative to the number of irrelevant clustering variables (e.g. if 7 out of 10 are relevant).  In these cases, the irrelevant clustering variables are discarded (i.e. strongly down-weighted), resulting in performance that is comparable with that achieved when the irrelevant clustering variables are artificially removed (e.g. consider $q = 7$ in Figure~\ref{fig03}).  However, when the number of relevant clustering variables is small relative to the number of irrelevant clustering variables (e.g. if only 3 out of 10 are relevant), variable selection can diminish performance;  as, in these cases, it is the relevant clustering variables that are discarded.    Since PReMiuM models the data as possessing only one (non-null) clustering structure, it tends to home in on a single ``dominant" clustering structure (here, the one that is defined by the majority of clustering variables).  Variable selection reinforces this by removing or down-weighting any clustering variables that define a different clustering structure, resulting in performance that is better if the dominant clustering also happens to be the relevant one, and worse if not.  

Crucially, we stress that the results seen here are not specific either to the inference or variable selection procedures implemented in PReMiuM, but are a consequence of failing to model multiple (non-null) views in the data, when they exist.    

\subsection{Breast cancer subtyping}\label{brcaexample}
In recent years, many authors \citep[e.g.][]{Shen2009,Kirk2012,Lock2013,Savage2013} have considered the problem of how to perform {\em integrative clustering}, using multiple 'omics datasets in order to better characterise cancer subtypes at the molecular level \citep[see also][for a reveiw]{Kristensen2014}.  Although multi-view approaches have not previously been applied for this purpose, they straightforwardly permit integrative clustering.  In the multi-view model, clustering variables are allocated to the same view if they define the same clustering structure.  Whether or not these clustering variables come from the same dataset or are of the same data type is irrelevant; all that matters is the clustering structure that they define.  

Here we apply semi-supervised multiview modelling to identify clusters among breast cancer tumour samples on the basis of reverse phase protein array (RPPA) and micro-RNA (miRNA) data from TCGA \citep{TCGA2012}.  A great deal of existing work has considered the use of mRNA expression data for identifying cancer subtypes, including the PAM50 predictive model for classifying breast cancer tumour samples on the basis of the expression of 50 genes \citep{Parker2009}.  We consider $n = 108$ breast cancer tumour samples, 66 of which have been classified on the basis of mRNA expression data as {\em basal-like} and 42 as {\em Luminal A}.  We use this classification as a binary response, to guide the clustering on the basis of the RPPA and miRNA data from TCGA.  The RPPA data comprise measurements on 171 proteins, while the miRNA data comprise measurements on 423 miRNAs.  

Before clustering, we process the miRNA and RPPA datasets as in \citet{Lock2013}.  To robustify against misspecification of models for continuous data, we take the additional pre-processing step of using tertile discretisation within each tumour sample, and treat the resulting data as categorical.  After pre-processing, the datasets are concatenated, so that our final working dataset has $p =  594$ clustering variables (corresponding to 171 proteins and 423 miRNAs).

\subsubsection{Results}\label{brcaResults}
\begin{figure}[!ht]\centering
    \begin{subfigure}[b]{0.31\textwidth}\centering
{\includegraphics[width=0.99\linewidth,trim = 0mm 0mm 0mm 0mm, clip]{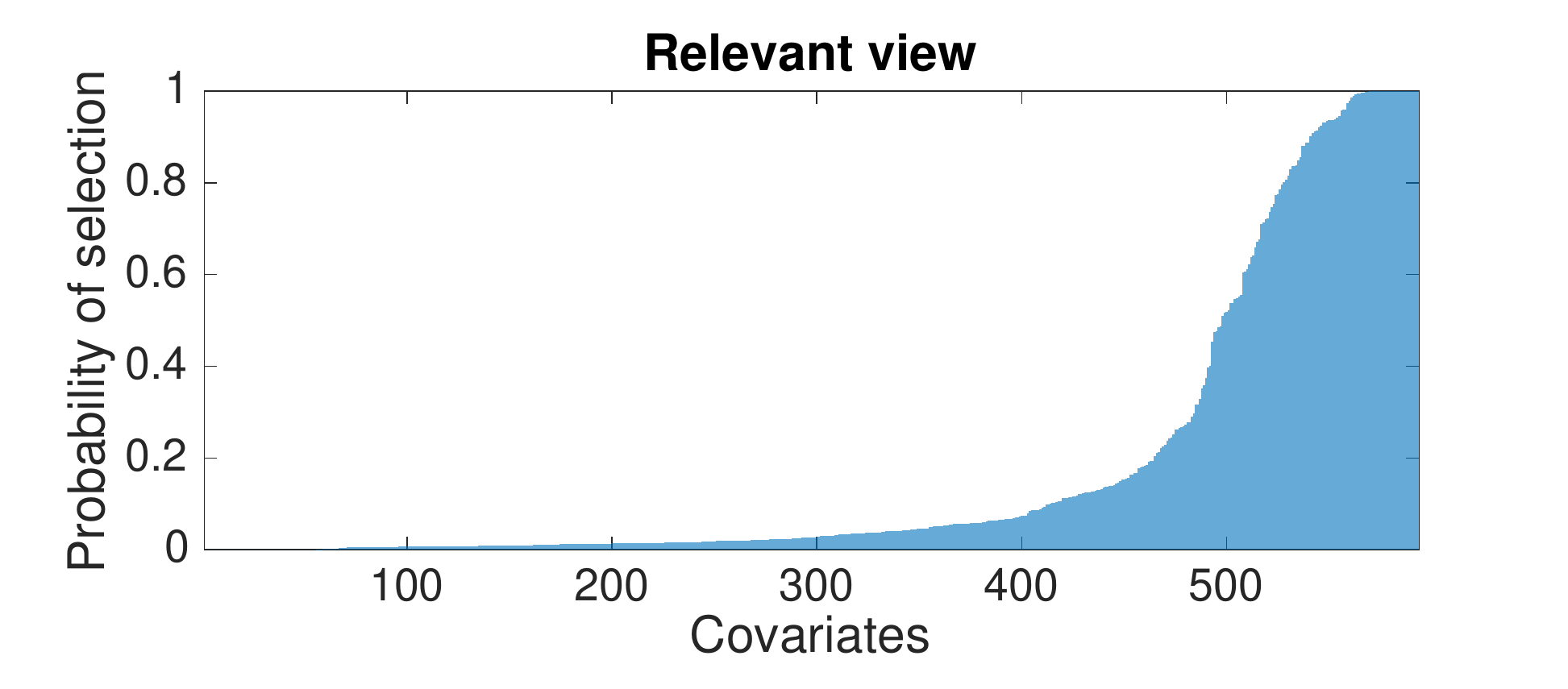}}
\caption{}
    \end{subfigure}
    ~
    \begin{subfigure}[b]{0.31\textwidth}\centering
{\includegraphics[width=0.99\linewidth,trim = 0mm 0mm 0mm 0mm, clip]{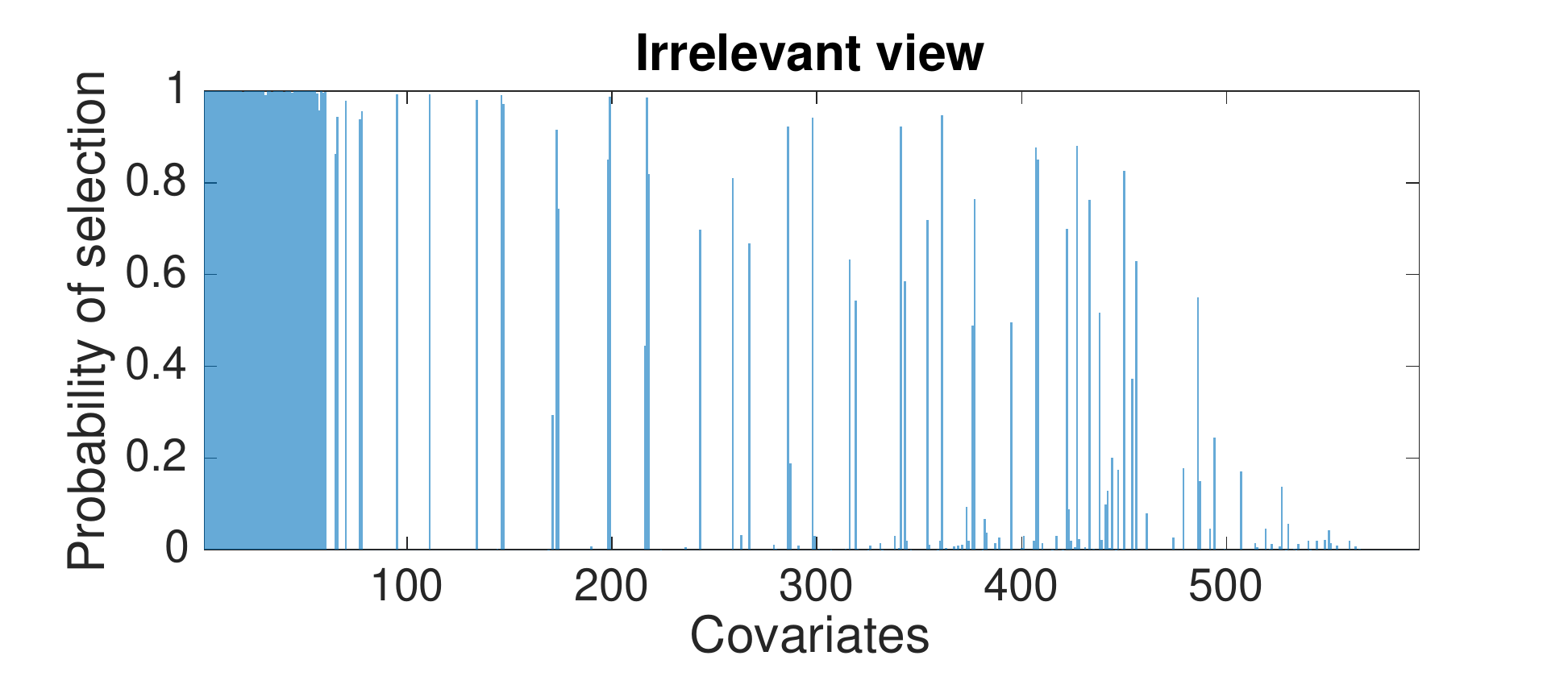}}
\caption{}
    \end{subfigure}
    ~
    \begin{subfigure}[b]{0.31\textwidth}\centering
{\includegraphics[width=0.99\linewidth,trim = 0mm 0mm 0mm 0mm, clip]{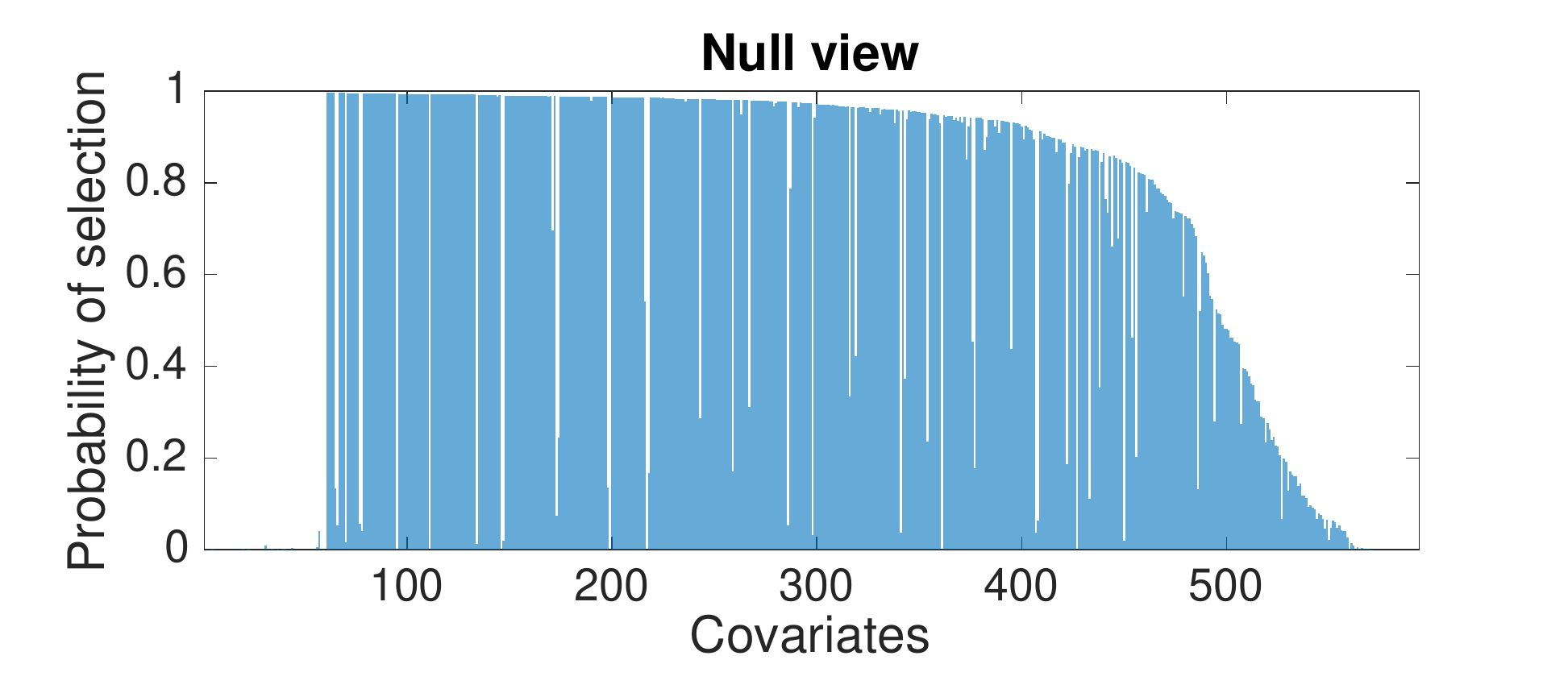}}
\caption{}
    \end{subfigure}
    \caption{Probability of each clustering variable being selected into the (a) relevant; (b) irrelevant; and (c) null views.  In each plot, the clustering variables are ordered  along the $x$-axis according to increasing probability of selection into the relevant view.}\label{selectionProbs}
\end{figure}

We fitted our semi-supervised multi-view model to the concatenated TCGA miRNA and RPPA data, assuming $L = 3$ views (1 relevant, 1 irrelevant, and 1 null view).  We performed 10,000 Gibbs sampling iterations, removing the first 5,000 as burn in, and then thinning to retain every 5-th draw.  For each clustering variable, we calculated Monte Carlo estimates of the probability of being selected into each view.  Bar plots of these probabilities are shown in Figure \ref{selectionProbs}.  The majority of clustering variables are selected with high probability into the null view, with relatively few selected with high probability into the other two views.      

\begin{figure}[!ht]\centering
    \begin{subfigure}[b]{0.49\textwidth}\centering
{\includegraphics[width=0.9\linewidth,trim = 0mm 0mm 0mm 0mm, clip]{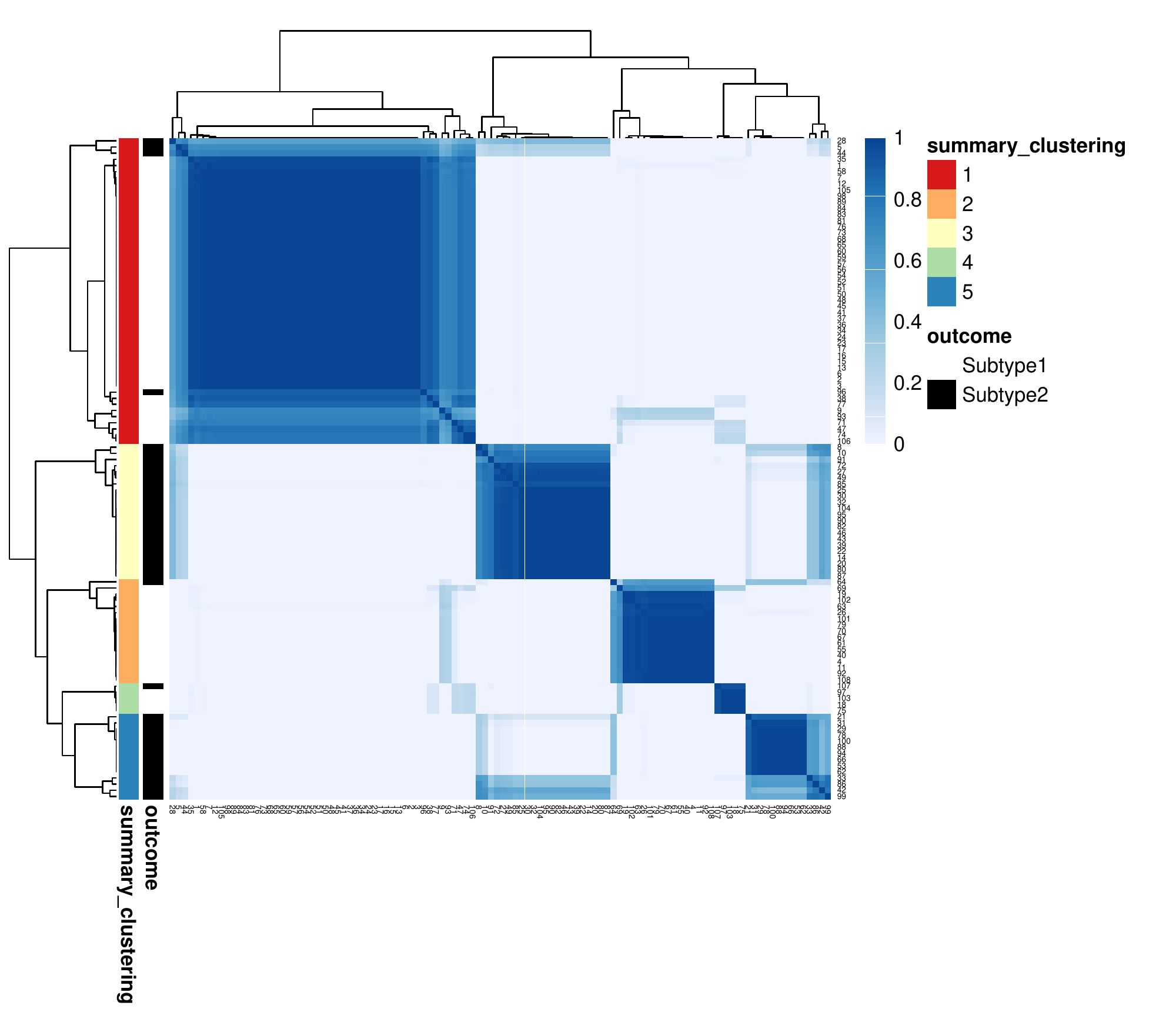}}
\caption{Posterior similarity matrix: relevant view}
    \end{subfigure}
    \begin{subfigure}[b]{0.49\textwidth}\centering
{\includegraphics[width=0.9\linewidth,trim = 0mm 0mm 0mm 0mm, clip]{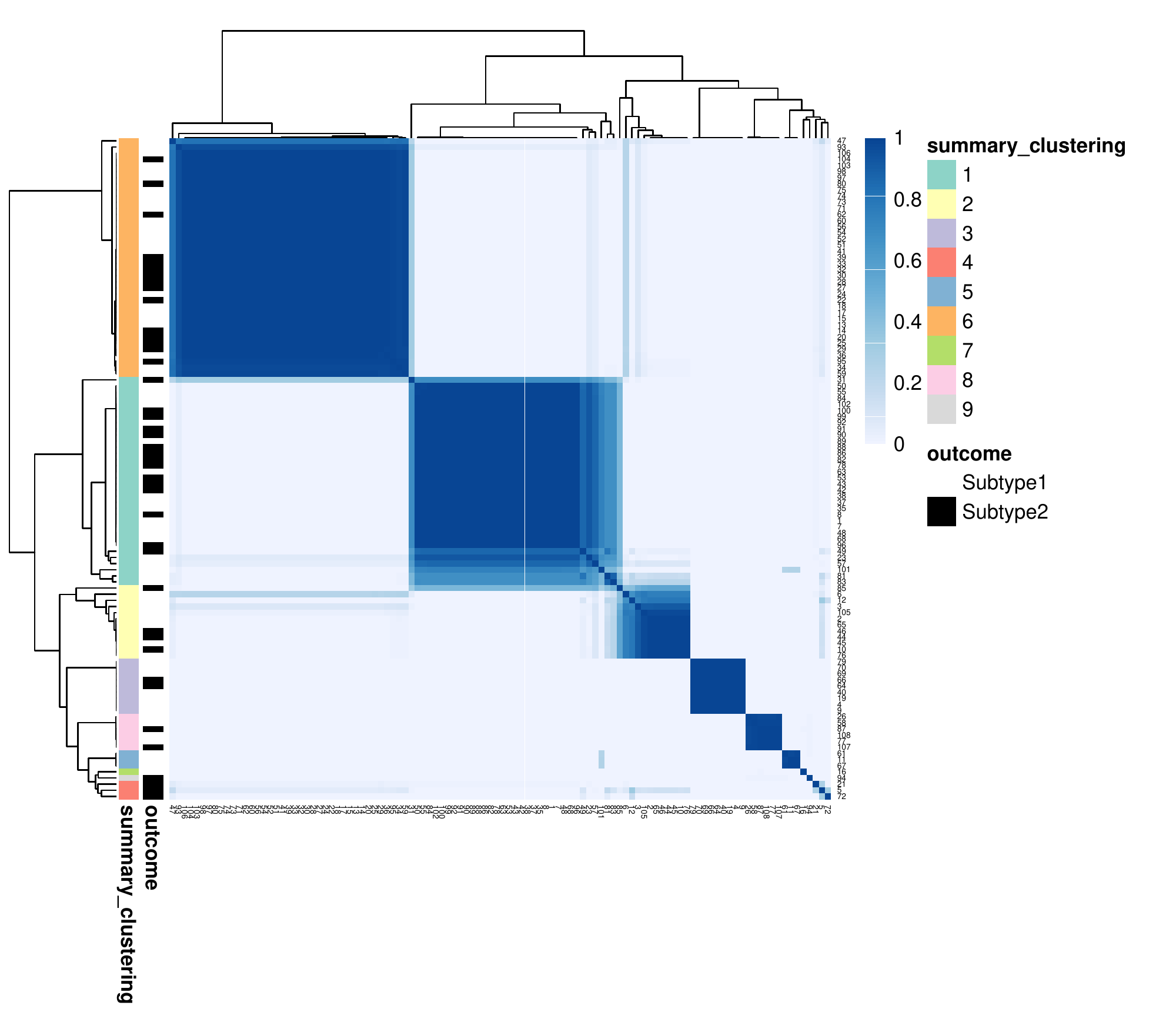}}
\caption{Posterior similarity matrix: irrelevant view}
    \end{subfigure}\caption{Posterior similarity matrices summarising the allocation of individuals to clusters in the (a) relevant; and (b) irrelevant views.  To aid visualisation, hierarchical clustering has been applied to both the rows and columns of the PSMs (as indicated by the dendrograms).}\label{PSMs}
\end{figure}

To summarise the clustering structure within each non-null view, we first calculated the {\em posterior similarity matrix} (PSM) for each view.  A PSM is an $n \times n$ matrix whose $i,j$-entry indicates the proportion of clusterings sampled from the posterior in which tumour sample $i$ and $j$ had the same cluster label \citep{Fritsch2009}.  Since PSMs summarise pairwise co-clustering probabilities, they provide a summary of the MCMC output that avoids challenges associated with label-switching.  The PSMs for the non-null views are provided in Figure~\ref{PSMs}.  It is clear from Figure~\ref{PSMs}a that, as desired, the clustering structure in the relevant view has a strong association with the subtype label.  As shown in Figure~\ref{PSMs}b, the irrelevant view also possesses a strong clustering structure, but one which is not associated with the subtype label.

\begin{figure}[!ht]\centering
    \begin{subfigure}[b]{0.49\textwidth}\centering
\includegraphics[height=6in]{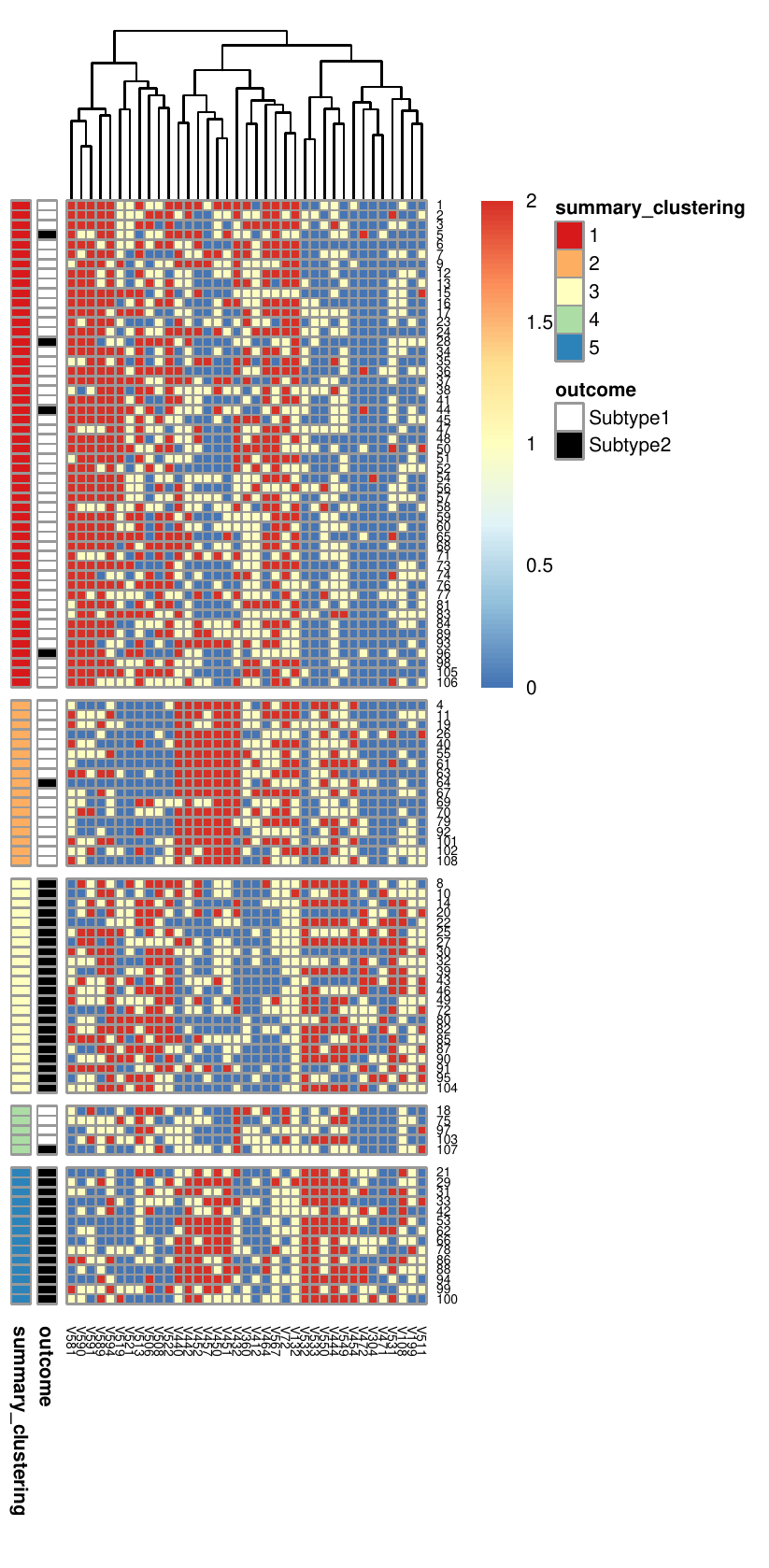}
\caption{Clustering structure: relevant view}
    \end{subfigure}
    \begin{subfigure}[b]{0.49\textwidth}\centering
\includegraphics[height=6in]{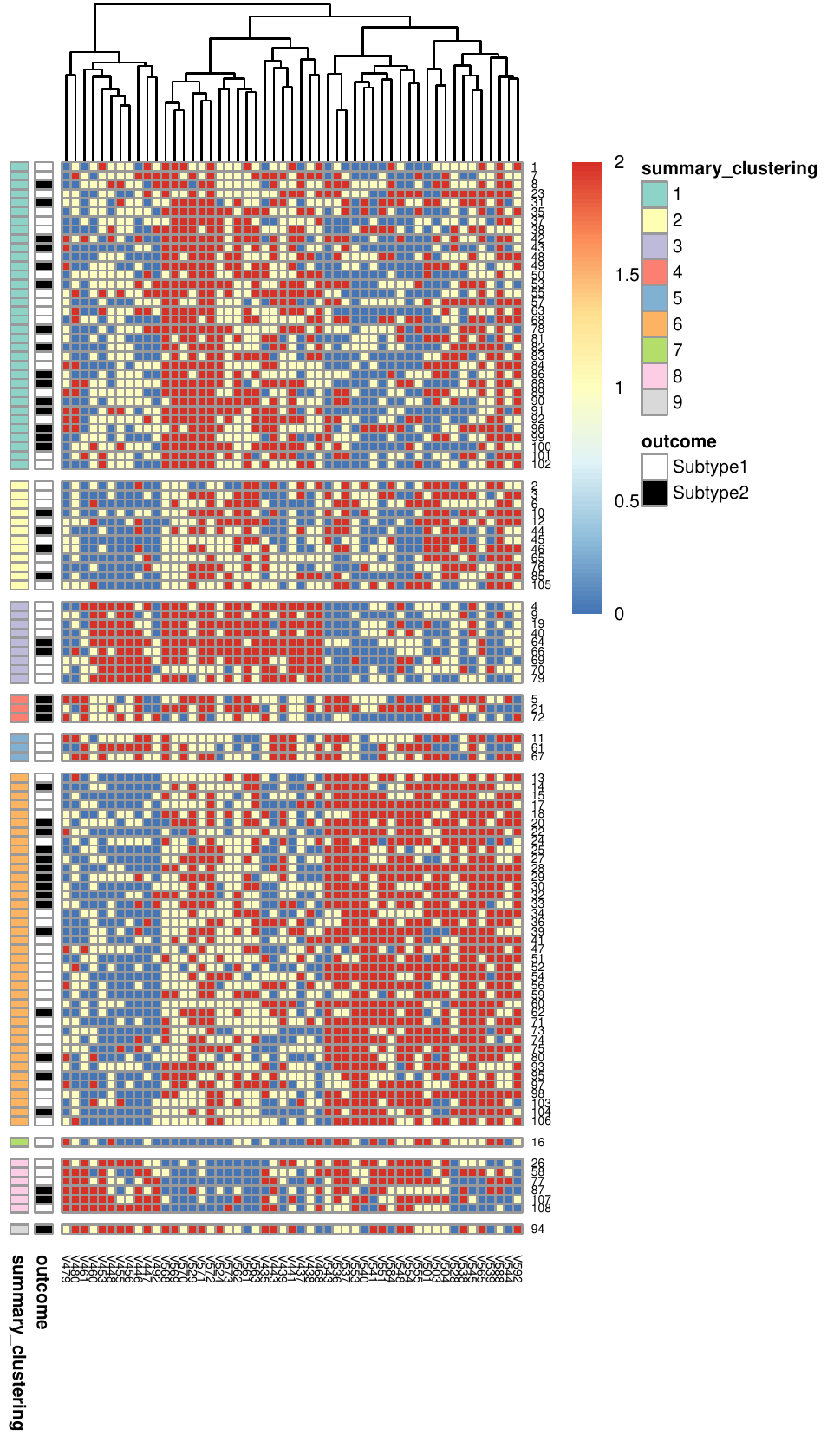}
\caption{Clustering structure: irrelevant view}
    \end{subfigure}\caption{Visualisation of the clustering structures in (a) relevant; and (b) irrelevant views.  Rows correspond to tumour samples, and the columns in each plot correspond to clustering variables selected with probability at least 0.90 of being selected into the (a) relevant; and (b) irrelevant views.  Hierarchical clustering has been applied to both the rows and columns of each plot, as indicated by the dendrograms.  The PAM50 subtypes for each tumour sample are indicated by an additional column shown to the left of each plot.  }\label{mv}
\end{figure}

To visualise the clusters in the relevant view, we thresholded the selection probabilities to retain only those clustering variables selected with probability at least 0.90 of being selected into the relevant view.  This left only 53 clustering variables, of which 32 were miRNAs and 21 were proteins.  A heatmap representation of the data for these 53 clustering variables is provided in Figure~\ref{mv}a, from which both the clustering structure and its association with the response (subtype) is clear.  We similarly visualised the clusters in the irrelevant view by retaining only those clustering variables selected with probability at least 0.90 of being selected into the irrelevant view.  This left 76 clustering variables, all of which were proteins.  The resulting heatmap representation is provided in Figure~\ref{mv}b.  In this case, while there is an evident clustering structure, it is not associated with tumour subtype. 

\section{Discussion}
We have demonstrated that, when there are multiple clustering structures present in data, existing (single view) clustering approaches can fail to recover the most relevant clustering structure, even when guided by an appropriate response (Section \ref{simresults}).  Moreover, traditional variable selection approaches for clustering do not necessarily improve matters, since they tend to select variables that define the {\em dominant} clustering structure, regardless of whether or not it is associated with a response of interest.  In Section \ref{brcaResults}, we have shown that real molecular datasets can and do possess multiple clustering structures, and that our semi-supervised multi-view model can allow both relevant and irrelevant structures to be identified.

While multi-view approaches clearly provide advantages relative to existing (single view) alternatives, computational considerations are a notable challenge.  Dirichlet process mixture models are already computationally costly, and the multi-view approach proposed here introduces an additional Dirichlet process mixture model for each additional view.  While restricting the number of views provided adequate results in the examples considered here, this need not be the case if the ``true" number of views in the data is greater than the selected value for $L$.  Both computational approaches \citep[such as parallelisation, as in][]{Suchard2010} and fast approximate inference procedures \citep[such as variational Bayes, as in][]{Guan2010} will be important considerations for future work.  

\section*{Funding}
PDWK and SR were supported by the MRC (MC\_UU\_00002/13 and MC\_UU\_00002/10) respectively.  FP and SR were supported by EPSRC project EP/R018561/1.

\section{Appendix}

\subsection{Unsupervised Bayesian mixture models}
We start by considering unsupervised mixture models of the following general form:
\begin{align}
p({\bf x}|\boldsymbol{\phi}, \boldsymbol{\pi}) & = \sum_{k=1}^K \pi_k f_{\bf x}({\bf x}|\boldsymbol{\phi}_k),
\end{align}
where $\boldsymbol{\pi} = [\pi_1, \ldots, \pi_K]$ is the vector of mixture weights and $\boldsymbol{\phi} = \{\boldsymbol{\phi}_1, \ldots, \boldsymbol{\phi}_K\}$ is the collection of all component-specific parameters.  We shall initially consider the case of finite $K$, and return later to the infinite mixture model. 

As is common for mixture models, we introduce latent component allocation variables, $z_i$, where $z_i = k$ if the $i$-th observation ${\bf x}_i $ is associated with the $k$-th component, and $p(z_i = k| \pi) = \pi_k$.  Then, 
\begin{align}
p({\bf x}_i|z_i, \boldsymbol{\phi}) = f_{\bf x}({\bf x}_i | \boldsymbol{\phi}_{z_i}),
\end{align}
and hence
\begin{align}
p({\bf x}_i,z_i = k| \boldsymbol{\phi}, \pi) &= f_{\bf x}({\bf x}_i | \boldsymbol{\phi}_{k})p(z_i = k| \boldsymbol{\pi})\\
&=f_{\bf x}({\bf x}_i | \boldsymbol{\phi}_{k})\pi_k.
\end{align}
Integrating out $z_i$ by summing over all $K$ possible values, we obtain (as we would hope):
\begin{align}
p({\bf x}_i| \boldsymbol{\phi}, \pi) &= \sum_{k = 1}^K \pi_k f_{\bf x}({\bf x}_i | \boldsymbol{\phi}_{k}).
\end{align}

Making the usual conditional independence assumptions, the full joint model for ${\bf x}_i, z_i, \boldsymbol{\phi}, \boldsymbol{\pi}$ is:
\begin{align}
p({\bf x}_i,z_i, \boldsymbol{\phi}, \pi) &= f_{\bf x}({\bf x}_i | \boldsymbol{\phi}_{z_i})p(z_i | \boldsymbol{\pi})p(\boldsymbol{\pi})p(\boldsymbol{\phi})\\
&= f_{\bf x}({\bf x}_i | \boldsymbol{\phi}_{z_i})p(z_i | \boldsymbol{\pi})p(\boldsymbol{\pi})\prod_{k = 1}^K p(\boldsymbol{\phi}_k),
\end{align}
where we assume independent priors for the component-specific parameters, $\phi_k$, and a symmetric Dirichlet prior for the mixture weights, $\pi_1, \ldots, \pi_K \sim \mbox{Dir}(\alpha/K)$.

For the full dataset, we have:
\begin{align}
p({\bf x}_1,\ldots, {\bf x}_n,z_1, \ldots, z_n, \boldsymbol{\phi}, \boldsymbol{\pi}) &= \left(\prod_{i=1}^n f_{\bf x}({\bf x}_i | \boldsymbol{\phi}_{z_i})p(z_i | \boldsymbol{\pi})\right)p(\boldsymbol{\pi})p(\boldsymbol{\phi})\\
&= \left(\prod_{i=1}^n f_{\bf x}({\bf x}_i | \boldsymbol{\phi}_{z_i})p(z_i | \boldsymbol{\pi})\right)p(\boldsymbol{\pi})\prod_{k = 1}^K p(\boldsymbol{\phi}_k).\label{joint}
\end{align}

\subsubsection{Inference via Gibbs sampling (finite $K$ case)}
Given Equation \eqref{joint}, it is straightforward to write down the conditionals for Gibbs sampling.  For the time being, we assume finite $K$ (from which we will later derive the infinite limit).  

\paragraph{Conditional for $\boldsymbol{\phi}_k$}
  By examination of the RHS of Equation \ref{joint}, we have:
\begin{equation}
p(\boldsymbol{\phi}_{k}| {\bf x}_1,\ldots, {\bf x}_n,z_1, \ldots, z_n, \boldsymbol{\phi}_{-k}, \pi_1, \ldots, \pi_K) \propto p(\boldsymbol{\phi}_k) \prod_{i: z_i = k} f_{\bf x}({\bf x}_i | \boldsymbol{\phi}_{z_i}),\label{postparm}
\end{equation} 
where $\boldsymbol{\phi}_{-k}$ denotes the set comprising all $\boldsymbol{\phi}_{j}$ for which $j\ne k$.  Thus the conditional for $\boldsymbol{\phi}_k$ is the posterior density for $\boldsymbol{\phi}_k$ given all ${\bf x}_i$ for which $z_i = k$.  If conjugate priors are taken, this posterior is available analytically, otherwise samples may be drawn by, for example, the Metropolis-Hastings algorithm.
Note that if there are no ${\bf x}_i$ for which $z_i = k$ (i.e. if the $k$-th component has no observations associated with it), then $\boldsymbol{\phi}_{k}$ is simply sampled from the prior, $p(\boldsymbol{\phi}_k)$.     

\paragraph{Conditional for $\pi$}
  By examination of the RHS of Equation \ref{joint}, we have:
\begin{align}
p(\pi_1, \ldots, \pi_K| {\bf x}_1,\ldots, {\bf x}_n,z_1, \ldots, z_n, \boldsymbol{\phi}_1, \ldots, \boldsymbol{\phi}_K) &\propto \left(\prod_{i=1}^n p(z_i | \pi_1, \ldots, \pi_K)\right)p(\pi_1, \ldots, \pi_K).
\end{align}
Hence, the conditional for $\pi$ is the posterior for $\pi$ given the values taken by the categorical latent allocation variables, $z_i$, $i = 1, \ldots, n$.  If we take a conjugate Dirichlet prior, this posterior is available in closed form.

\paragraph{Conditional for $z_i$}
  By examination of the RHS of Equation \ref{joint}, we have:
\begin{align}
p(z_i = k| {\bf x}_1,\ldots, {\bf x}_n,\boldsymbol{\phi}_1, \ldots, \boldsymbol{\phi}_K, \pi_1, \ldots, \pi_K, z_{-i}) &\propto p(z_i=k|\pi) f_{\bf x}({\bf x}_i | \phi_{k}),\label{cicond1}\\
&= \pi_k  f_{\bf x}({\bf x}_i | \phi_{k}),
\end{align}
where $z_{-i}$ denotes the set comprising all $z_{j}$ for which $j\ne i$.  Since $\sum_{k = 1}^K p(z_i = k| {\bf x}_1,\ldots, {\bf x}_n,\boldsymbol{\phi}, \pi, z_{-i}) = 1$, it follows that the conditional is:  
\begin{align}
p(z_i = k| {\bf x}_1,\ldots, {\bf x}_n,\boldsymbol{\phi}, \pi, z_{-i}) &= \frac{\pi_k  f_{\bf x}({\bf x}_i | \phi_{k})}{\sum_{k=1}^K \pi_k  f_{\bf x}({\bf x}_i | \phi_{k})},
\end{align}
which may be straightforwardly evaluated for finite $K$.
\paragraph{Marginalising $\pi$}
Taking a conjugate Dirichlet prior for $\pi$, an alternative strategy is to marginalise $\boldsymbol{\pi}$ rather than to sample it.  We assume a symmetirc Dirichlet prior with concentration parameter $\alpha/K$.  

{\bf Note} that the $z_i$'s are only conditionally independent of one another given $\boldsymbol{\pi}$, so if we marginalise $\boldsymbol{\pi}$ then we must be careful to model the dependence of $z_i$ on $z_{-i}$ in our conditional for~$z_i$. 

We have    
\begin{align}
p(z_i = k| {\bf x}_1,\ldots, {\bf x}_n,\boldsymbol{\phi}, \boldsymbol{\pi}, z_{-i}, \alpha) &\propto p(z_i=k|z_{-i}, \pi, \alpha) f_{\bf x}({\bf x}_i | \phi_k) \mbox{\quad\quad\quad [cf. Equation \eqref{cicond1}]}.
\end{align}
To marginalise $\boldsymbol{\pi}$, we must therefore evaluate $\int_\pi p(z_i=k|z_{-i}, \boldsymbol{\pi}, \alpha) p(\boldsymbol{\pi}|\alpha)d\boldsymbol{\pi} = p(z_i=k|z_{-i}, \alpha)$, which is the conditional prior for $z_i$ given the values for the other latent allocation variables, $z_{-i}$.  

We have,
\begin{align}
p(z_i=k|z_{-i}, \alpha) &= \frac{p(z_i=k,z_{-i}|  \alpha)}{p(z_{-i}|  \alpha)} \\
&= \frac{\int_{\boldsymbol{\pi}} p(z_i=k,z_{-i}| \boldsymbol{\pi}, \alpha)p(\boldsymbol{\pi}|\alpha)d\boldsymbol{\pi}}{ \int_{\boldsymbol{\pi}} p(z_{-i}| \boldsymbol{\pi}, \alpha)p(\boldsymbol{\pi}|\alpha)d\boldsymbol{\pi}}\\
&= \frac{\int_{\boldsymbol{\pi}} p(z_i=k,z_{-i}| \boldsymbol{\pi})p(\boldsymbol{\pi}|\alpha)d\boldsymbol{\pi}}{ \int_{\boldsymbol{\pi}} p(z_{-i}| \boldsymbol{\pi})p(\boldsymbol{\pi}|\alpha)d\boldsymbol{\pi}}, \label{cicond2}  
\end{align}
where in the final line we exploit the fact that the $z_i$'s are conditionally independent of $\alpha$, given~$\pi$.

In order to proceed, we must evaluate this fraction.  To do this we require a standard result about Dirichlet distributions, which says that moments of random variables distributed according to a symmetric Dirichlet distribution with parameter $\alpha/K$ can be expressed as follows:
\begin{equation}
E\left[\prod_{k = 1}^K \pi_k^{m_k} \right] = \frac{\Gamma(\sum_{k=1}^K (\alpha/K))}{\Gamma(\sum_{k=1}^K ((\alpha/K) + m_k))}\times \prod_{k=1}^K \frac{\Gamma ((\alpha/K) + m_k)}{\Gamma(\alpha/K)},\label{standard}
\end{equation}
where the $m_k$'s are any natural numbers. 

Moreover, we note the following two equalities: 
$$p(z_i = k, z_{-i}| \pi) = \pi_k^{n_{-i,k} + 1}\prod_{\substack{c= 1,\ldots,K\\
                  c \ne k}} \pi_c^{n_{-i,c}}, $$
                  and
$$p(z_{-i}| \pi) = \pi_k^{n_{-i,k}}\prod_{\substack{c= 1,\ldots,K\\
                  c \ne k}} \pi_c^{n_{-i,c}}, $$
where $n_{-i,c}$ is the number of $z_j$'s with $j \ne i$ for which $z_j = c$.  It then follows that we may use the result given in Equation \eqref{standard} in order to evaluate the numerator and denominator in the RHS of Equation \eqref{cicond2}.  After some algebra, and exploiting the property of Gamma functions that $\Gamma(t+1) = t\Gamma(t)$, we obtain:
\begin{align}
p(z_i=k|z_{-i}, \alpha) &= \frac{n_{-i,k} + \alpha/K}{n - 1 + \alpha}. \label{cicond3}  
\end{align}
Hence,
\begin{align}
p(z_i = k| {\bf x}_1,\ldots, {\bf x}_n,\boldsymbol{\phi}, z_{-i}, \alpha) &\propto  \frac{n_{-i,k} + \alpha/K}{n - 1 + \alpha} \times f_{\bf x}({\bf x}_i | \phi_k).
\end{align}
Moreover, since $K$ is finite, we may straightforwardly evaluate the equality:
\begin{align}
p(z_i = k| {\bf x}_1,\ldots, {\bf x}_n,\boldsymbol{\phi}, z_{-i}, \alpha) &= \frac{1}{Z} \frac{n_{-i,k} + \alpha/K}{n - 1 + \alpha} \times f_{\bf x}({\bf x}_i | \phi_k),\label{finite_nomarg}
\end{align} 
where
\begin{equation}
Z = \sum_{c =1}^K \left(\frac{n_{-i,c} + \alpha/K}{n - 1 + \alpha} \times f_{\bf x}({\bf x}_i | \phi_c)\right).  
\end{equation}

\paragraph{Marginalising $\phi$}
Similarly, if a conjugate prior is available for the $\phi_k$'s, then these may be marginalised too.  {\bf Note} that (similar to the case with the $z_i$'s when we marginalised $\pi$) the ${\bf x}_i$'s are only conditionally independent of one another given the $\phi_k$'s and the $z_i$'s, so if we marginalise $\phi$ then we must be careful to model the dependence of $v_i$ on $v_{-i}$ in our conditional for $z_i$.  

After some algebra, it is straightforward to show that marginalising $\phi$ gives the following for the conditional for~$z_i$:
\begin{align}
p(z_i = k| {\bf x}_1,\ldots, {\bf x}_n,z_{-i}, \alpha) &= \frac{1}{Z} \frac{n_{-i,k} + \alpha/K}{n - 1 + \alpha} \times \int_{\phi_k}f_{\bf x}({\bf x}_i | \phi_k) p(\phi_k|{\bf x}_{-i,k}) d\phi_k,\label{condci4}
\end{align} 
where ${\bf x}_{-i,k}$ denotes all observations ${\bf x}_j$ for which $j \ne i$ and $z_j = k$, and hence $p(\phi_k|{\bf x}_{-i,k})$ is the posterior for $\phi_k$ given all of the observations currently associated with component $k$ (excluding ${\bf x}_i$).  If there are no ${\bf x}_j$ for which $j \ne i$ and $z_j = k$ (i.e. if $k$ is a component to which no other observations have been allocated), then we say that the $k$-th component is {\em empty} and define $p(\phi_k|{\bf x}_{-i,k}):= p(\phi_k)$ to be the prior for $\phi_k$.  

When implementing the sampler, it is useful to observe that $$p(\phi_k|{\bf x}_{-i,k}) = \frac{f_{\bf x}({\bf x}_{-i,k}|\phi_k) p(\phi_k)}{\int_{\phi_k}f_{\bf x}({\bf x}_{-i,k}|\phi_k) p(\phi_k)d\phi_k}, \mbox{ if the $k$-th component is not empty.}$$  
Hence, still assuming that the $k$-th component is not empty, the integral in Equation \eqref{condci4} is
\begin{equation}
\int_{\phi_k}f_{\bf x}({\bf x}_i | \phi_c) p(\phi_k|{\bf x}_{-i,k}) d\phi_k = 
\frac{\int_{\phi_k}f_{\bf x}({\bf x}_i,{\bf x}_{-i,k}|\phi_k) p(\phi_k)d\phi_k}{\int_{\phi_k}f_{\bf x}({\bf x}_{-i,k}|\phi_k) p(\phi_k)d\phi_k},\label{marg}
\end{equation}
which is a ratio of marginal likelihoods: one in which we include ${\bf x}_i$ amongst the observations associated with component $k$, and one in which we exclude ${\bf x}_i$ from the observations associated with component $k$.  

This expression aids the interpretation of the sampler: at each iteration, and for each component, we weigh the evidence that ${\bf x}_i$ is associated with component $k$ against the evidence that ${\bf x}_i$ is {\em not} associated with component $k$ (given the other observations currently associated with that component, ${\bf x}_{-i,k}$).  Intuitively, this expression ensures that we are more likely to allocate ${\bf x}_i$ to a component to which similar observations have previously been allocated.

Note also that the term $\frac{n_{-i,k} + \alpha/K}{n - 1 + \alpha} $ in Equation \eqref{condci4} represents the conditional prior probability that ${\bf x}_i$ should be allocated to component $k$ represents our prior belief that we should allocate ${\bf x}_i$ to component $k$, given the allocation of all of the other observations.  Since $n_{-i,k}$ is in the numerator, this expresses a ``rich-get-richer" prior belief; i.e. that, {\em a priori}, we are more likely to assign ${\bf x}_i$ to a component that already has many observations assigned to it, rather than to one with fewer.    

\paragraph{Final note}
Note that, having marginalised the $\pi_k$'s and $\phi_k$'s, we may use Equation \eqref{condci4} to sample just the $z_i$'s, without having to sample any other parameters.  The one exception is the $\alpha$ hyperparameter, which we may either fix or sample (using, for example, the approach described in Escobar and West, 1995).  

\subsubsection{Conditionals for Gibbs sampling ($K$ infinite)}
The infinite case can be derived from the finite case by considering the limit $K \rightarrow \infty$.  The key change is to the conditional priors for the component allocation variables, $p(z_i = k|z_{-i}, \alpha )$.  There are two cases to consider: 

\paragraph{1. $k$ is the label for an existing (non-empty) component:}
This is the case where $k = z_j$ for some $j \ne i$.  By considering the limit as $K\rightarrow \infty$ of Equation \eqref{cicond3}, we obtain the following:

\begin{align}
\mbox{If $k = z_j$ for some $j \ne i$, then\qquad} p(z_i=k|z_{-i}, \alpha) &= \frac{n_{-i,k}}{n - 1 + \alpha}. \label{infcicond}  
\end{align}

\paragraph{2. $k$ is a new label:}  This is the case where $k \ne z_j$ for all $j \ne i$.  Hence, the probability we seek is $p(z_i \ne z_j \mbox{ for all } j \ne i| z_{-i}, \alpha)$.  Calculation of this probability is aided by using the following identity: 
\begin{equation}
1 = p(z_i \ne z_j \mbox{ for all } j \ne i| z_{-i}, \alpha) + \sum_{k \in \mathcal{C}_{-i}} p(z_i=k|z_{-i}, \alpha), \notag
\end{equation}
where $\mathcal{C} _{-i}= \{k:  z_j = k \mbox{ for some } j\ne i \}$ is the set of all current component labels (excluding $z_i$).  Using this identity, we obtain (after a little algebra) the following: 
\begin{equation}
p(z_i \ne z_j \mbox{ for all } j \ne i| z_{-i}, \alpha) = \frac{\alpha}{n - 1 + \alpha}.\label{infcicond2}
\end{equation}
 
 \paragraph{Conditional for $z_i$}
Using Equations \eqref{infcicond} and \eqref{infcicond2}, it is straightforward to show (cf. Equation \eqref{condci4}, and making use of Equation \eqref{marg}) that we have the following conditionals for $z_i$:
\begin{align}
\hspace{-0.7cm}\mbox{If $k = z_j$ for some $j \ne i$, then\qquad} p(z_i=k|{\bf x}_1,\ldots, {\bf x}_n,z_{-i}, \alpha) &= b\frac{n_{-i,k}}{n - 1 + \alpha} \frac{\int_{\phi_k}f_{\bf x}({\bf x}_i,{\bf x}_{-i,k}|\phi_k) p(\phi_k)d\phi_k}{\int_{\phi_k}f_{\bf x}({\bf x}_{-i,k}|\phi_k) p(\phi_k)d\phi_k}, \label{infcond1}\\
\mbox{and \qquad} p(z_i \ne z_j \mbox{ for all } j \ne i| {\bf x}_1,\ldots, {\bf x}_n, z_{-i}, \alpha) &= b\frac{\alpha}{n - 1 + \alpha} \int_{\phi_k}f_{\bf x}({\bf x}_i | \phi_k) p(\phi_k) d\phi_k,\label{infcond2}
\end{align}
where $b$ is a normalising constant that ensures that
\begin{equation}
p(z_i \ne z_j \mbox{ for all } j \ne i| {\bf x}_1,\ldots, {\bf x}_n, z_{-i}, \alpha) + \sum_{k \in \mathcal{C}_{-i}} p(z_i=k|{\bf x}_1,\ldots, {\bf x}_n,z_{-i}, \alpha) = 1. \notag
\end{equation}

Up to changes in notation and expansion of some terms, these expressions are identical to those given in Equation~(3.7) of Neal (2000).

\subsection{Profile regression for categorical data}\label{profileRegression}
Recall that our data comprises input-output (i.e. clustering variable vector-response) pairs, $D = \{{\bf v}_i \}_{i = 1}^n$, where ${\bf v}_i = [{\bf x}_i, y_i]^\top$, with ${\bf x}_i$  and $y_i$ denoting the $i$-th clustering variable vector and response, respectively.  We now consider the case in which the clustering variables and response are all categorical.  

\subsubsection{Modelling the clustering variables}
We now assume that each clustering variable (i.e. each element of the vector ${\bf x}_i$) is categorical, with the $j$-th clustering variable having $R_j$ categories, which we label as $1, 2, \ldots, R_j$.  We model the data using categorical distributions.  We define $\phi_{k,j,r}$ to be the probability that the $j$-th clustering variable takes value $r$ in the $k$-th component, and write $\Phi_{k,j} = [\phi_{k,j,1}, \phi_{k,j,2}, \ldots, \phi_{k,j,R_j} ]$ for the collection of probabilities associated with the $j$-th clustering variable in the $k$-th component.  We further define $\Phi_{k} = \{\Phi_{k,1}, \Phi_{k,2}, \ldots, \Phi_{k,J}\}$ to be the collection of all probabilities (over all $J$ clustering variables) associated with the $k$-th component, and $\Phi = \{ \Phi_{k} \}_{k\in \mathcal{C}}$ to be the set of all $\Phi_{k}$'s that are associated with non-empty components (here, $\mathcal{C} = \{k: z_i = k \mbox{ for some } i \in \{1,\ldots,n\}\}$).  

We assume that the clustering variables are conditionally independent, given their component allocation, so that 
\begin{align}      
 f_{\bf x}({\bf x}_i = [x_{i1}, x_{i2}, \ldots, x_{iJ}]|\boldsymbol{\Phi}, z_i = k) &= \phi_{k,j,x_{i1}} \phi_{k,j,x_{i2}} \ldots \phi_{k,j,x_{iJ}}\\
 &= \prod_{j = 1}^J  \phi_{k,j,x_j}\label{likeliprod}
\end{align}

\paragraph{Conditional for $\Phi_{k,j}$}
From Equation \eqref{postparm}, the conditional that we require for Gibbs sampling is the posterior for $\Phi_{k,j}$, given the observations associated with the $k$-th component.  For each $j$, we adopt a conjugate Dirichlet prior for $\Phi_{k,j}$, $$\Phi_{k,j} \sim \mbox{Dirichlet}({\bf a}_j),$$ where ${\bf a}_j = [a_{j,1}, \ldots, a_{j,R_j}]$ is the vector of concentration parameters.
The posterior is then:
\begin{align}
\Phi_{k,j}|{ x}_{i_1,j}, { x}_{i_2,j}, \ldots, { x}_{i_{n_k},j}, {\bf a}_j \sim \mbox{Dirichlet}({\bf a}_j + [s_{k,j,1}, s_{k,j,2}, \ldots, s_{k,j,R_j}]),
\end{align}
where ${\bf x}_{i_1}, {\bf x}_{i_2}, \ldots, {\bf x}_{i_{n_k}}$ are the observations associated with component $k$, and $s_{k,j,r}$ is defined to be the number of observations associated with component $k$ for which the $j$-th clustering variable is in category $r$.  

\paragraph{Marginalising $\Phi_{k,j}$}\label{marginphi}
We may also integrate out $\Phi_{k,j}$ in order to write down the marginal likelihood associated with ${ x}_{i_1,j}, { x}_{i_2,j}, \ldots, { x}_{i_{n_k},j}$.  Note that the marginal likelihood is (by definition) the prior expectation of the product $\phi_{k,j,1}^{s_{k,j,1}}\ldots\phi_{k,j,{R_j}}^{s_{k,j,{R_j}}}$.  We may therefore use the same standard result that was used to derive Equation \eqref{standard} in order to immediately write down the marginal likelihood.  Still assuming that ${\bf x}_{i_1}, {\bf x}_{i_2}, \ldots, {\bf x}_{i_{n_k}}$ are the observations associated  with component $k$, we have:
\begin{equation}
p({ x}_{i_1,j}, { x}_{i_2,j}, \ldots, { x}_{i_{n_k},j}| {\bf a}_j) = \frac{\Gamma(\sum_{r=1}^{R_j}a_{j,r} )}{\Gamma(\sum_{r=1}^{R_j}(a_{j,r}+ s_{k,j,r}) )}\times \prod_{r = 1}^{R_j} \frac{\Gamma(a_{j,r} + s_{k,j,r})}{\Gamma(a_{j,r})}.\label{catmarg}
\end{equation}
To shorten notation, define $X_{k} = \{ {\bf x}_{i_1}, {\bf x}_{i_2}, \ldots, {\bf x}_{i_{n_k}} \}$ to be the set of observations associated with component $k$, and $X_{k, j} = \{ { x}_{i_1,j}, { x}_{i_2,j}, \ldots, { x}_{i_{n_k},j} \}$ to be the set containing the $j$-th elements of the vectors in $X_{k}$.  Since we assume that the clustering variables are conditionally independent, given their component allocation, it follows that:
    
\begin{equation}
p(X_{k}| {\bf a}_1, \ldots, {\bf a}_J) = \prod_{j = 1}^J p(X_{k,j}|{\bf a}_j),\label{marglikelix}
\end{equation}
where $p(X_{k,j}|{\bf a}_j) = p({ x}_{i_1,j}, { x}_{i_2,j}, \ldots, { x}_{i_{n_k},j}| {\bf a}_j)$ is as given in Equation \eqref{catmarg}.

\subsubsection{Modelling the response}

We assume that the response, $y_i$, is also categorical (this time with $R_y$ categories), and model using a categorical distribution.  Similar to before, we define $\theta_{k,r}$ to be the probability that $y$ takes value $r$ in the $k$-th component, and write $\Theta_k = [\theta_{k,1}, \theta_{k,1}, \ldots, \theta_{k,{R_y}}]$ to be the collection of all probabilities associated with the $k$-th component.  Also define $Y_k = \{y_{i_1}, y_{i_2}, \ldots, y_{i_{n_k}} \}$ to be the set of $y$'s allocated to component $k$.  
\paragraph{Conditional for $\theta_{k}$}
We adopt a conjugate Dirichlet prior for $\theta_k$, 
$$\theta_{k} \sim \mbox{Dirichlet}({\bf a}_y),$$ where ${\bf a}_y = [a_{y,1}, \ldots, a_{j,R_y}]$ is the vector of concentration parameters.  Similar to previously, the posterior for $\theta_{k}$ is then
\begin{align}
\theta_{k}|Y_k, {\bf a}_y \sim \mbox{Dirichlet}({\bf a}_y + [s_{k,y,1}, s_{k,y,2}, \ldots, s_{k,y,R_y}]),
\end{align}       
where $s_{k,y,r}$ is the number of observations in $Y_k$ that are in the $r$-th category.  

\paragraph{Marginalising $\theta_{k}$}\label{margintheta}
As before, we may also calculate the marginal likelihood, $p(Y_k| {\bf a}_j)$.  We have       
\begin{equation}
p(Y_k| {\bf a}_y) = \frac{\Gamma(\sum_{r=1}^{R_y}a_{y,r} )}{\Gamma(\sum_{r=1}^{R_y}(a_{y,r}+ s_{k,y,r}) )}\times \prod_{r = 1}^{R_y} \frac{\Gamma(a_{y,r} + s_{k,y,r})}{\Gamma(a_{y,r})}.\label{catrespmarg}
\end{equation}
\subsubsection{Joint marginal likelihood}
In order to proceed, we need an expression for the marginal likelihood associated with $V_k = \{X_k, Y_k\}$.  We assume that $y$ and ${\bf x}$ are conditionally independent, given their component allocation.  It follows that
\begin{equation}
p(V_{k}| {\bf a}_1, \ldots, {\bf a}_J, {\bf a}_y) = p(Y_k|{\bf a}_y)\prod_{j = 1}^J p(X_{k,j}|{\bf a}_j),
\end{equation}
where the expressions for $p(Y_k|{\bf a}_y)$ and $p(X_{k,j}|{\bf a}_j)$ are as given previously.  Note that:
\begin{enumerate}
  \item Setting $V_k = \{{\bf x}_i, {\bf x}_{-i,k}\}$, we can evaluate the marginal likelihood in the numerator in Equation \eqref{infcond1}.
  \item Setting $V_k = \{{\bf x}_{-i,k}\}$, we can evaluate the marginal likelihood in the denominator in Equation \eqref{infcond1}.
  \item Setting $V_k = \{{\bf x}_i\}$, we can evaluate the marginal likelihood in Equation \eqref{infcond2}.  
  \end{enumerate}

Thus, we may evaluate all of the terms required for the conditionals for $z_i$, and hence (leaving aside, for the time being, the problem of sampling $\alpha$) we have everything we need in order to perform inference for our model.  

\subsection{Variable selection}\label{variableSelection}
\subsubsection{A null model for the clustering variable data}
We introduce a null model for the ${\bf x}$ data, under the assumption that there is no clustering structure.  Under the null model, we again model the clustering variable data using categorical distributions, and -- similar to previously -- define $\Phi_{0,j} = [\phi_{0,j,1}, \phi_{0,j,2}, \ldots, \phi_{0,j,R_j} ]$ and $\Phi_{0} = \{\Phi_{0,1}, \Phi_{0,2}, \ldots, \Phi_{0,J}\}$.  However, under the null model, we assume that there is no clustering structure (or, equivalently, that the data form a single large cluster).  The likelihood associated with ${\bf x}_i$ therefore does not involve a component allocation variable, $z_i$, and is instead:
\begin{align}      
 f_{\bf x}({\bf x}_i = [x_{i1}, x_{i2}, \ldots, x_{iJ}]^\top|\boldsymbol{\Phi}_0)
 = \prod_{j = 1}^J  \phi_{0,j,x_{ij}}\label{likeliprodnull}
\end{align}
In our null model, the observations ${\bf x}_1, \ldots, {\bf x}_n$ are assumed to be conditionally independent given $\boldsymbol{\Phi}_0$.  It follows that, under the null model, the likelihood associated with the full clustering variable dataset is:
\begin{align}      
 f_{\bf x}({\bf x}_1, \ldots, {\bf x}_n|\boldsymbol{\Phi}_0) = \prod_{i = 1}^n \prod_{j = 1}^J  \phi_{0,j,x_{ij}}.\label{likeliprodnullfull}
\end{align}

\subsubsection{Variable selection}
We consider the possibility that only some of the clustering variables are relevant for determining the clustering structure, while the others are irrelevant.  We introduce binary indicators $\gamma_j$, such that $\gamma_j$ = 1 means that the $j$-th clustering variable is relevant for the clustering structure, while $\gamma_j$ = 0 means that the variable is irrelevant.  

Informally, we wish to use the full mixture model for the relevant variables, and the null model for the irrelevant variables. More precisely, we consider the following model for ${\bf x}$:
\begin{align}      
 f_{\bf x}({\bf x}_i = [x_{i1}, x_{i2}, \ldots, x_{iJ}]^\top|\boldsymbol{\Phi}, \boldsymbol{\Phi}_0, z_ i = k)
 &= \prod_{j = 1}^J  \phi_{k,j,x_{ij}}^{\mathbb{I}(\gamma_j=1)}\phi_{0,j,x_{ij}}^{\mathbb{I}(\gamma_j=0)}.\label{likeliprod2}
\end{align}

\subsubsection{Inference}
We consider how the introduction of the $\gamma_j$'s affects how we infer the model parameters.  As previously, we start by writing down the full joint model:

\begin{equation}
\begin{split}
p({\bf x}_1,\ldots, {\bf x}_n,&z_1, \ldots, z_n, \boldsymbol{\theta}, \boldsymbol{\Phi}, \boldsymbol{\Phi}_0, \pi, \alpha, \gamma) =\\ &p(\pi|\alpha)p(\Phi_0)\left(\prod_{j=1}^Jp(\gamma_j)\right)\left(\prod_{k = 1}^K  p(\Phi_k)p(\theta_k) \right)\left(\prod_{i=1}^n   p(z_i | \pi) \theta_{z_i, y_i} \left(\prod_{j=1}^J  \phi_{z_i,j,x_{ij}}^{\mathbb{I}(\gamma_j=1)}\phi_{0,j,x_{ij}}^{\mathbb{I}(\gamma_j=0)} \right)\right).\label{fsjoint}
\end{split}
\end{equation}

\paragraph{Inferring $\Phi_{k,j}$}
As before, we adopt independent Dirichlet priors for $\Phi_{k,j}$, 
\begin{equation}
\Phi_{k,j} \sim \mbox{Dirichlet}({\bf a}_j),\label{prior1}
\end{equation}
where ${\bf a}_j = [a_{j,1}, \ldots, a_{j,R_j}]$ is the vector of concentration parameters.  

It is clear from Equation \eqref{fsjoint} that the conditional for $\Phi_{k,j}$ is
\begin{equation}
p(\Phi_{k,j}|{\bf x}_1,\ldots, {\bf x}_n,z_1, \ldots, z_n, \boldsymbol{\theta}, \boldsymbol{\Phi}_{-{k,j}}, \boldsymbol{\Phi}_0, \pi, \alpha, \gamma) \propto p(\Phi_{k,j}) \prod_{i=1}^n  \prod_{j=1}^J  \phi_{z_i,j,x_{ij}}^{\mathbb{I}(\gamma_j=1)}.
\end{equation}

If $\gamma_j = 1$, then -- as before -- the conditional for $\Phi_{k,j}$ is the posterior for $\Phi_{k,j}$ given all observations currently associated with component $k$, i.e. 
\begin{align}
\Phi_{k,j}|\gamma_j = 1,{ x}_{i_1,j}, { x}_{i_2,j}, \ldots, { x}_{i_{n_k},j}, {\bf a}_j \sim \mbox{Dirichlet}({\bf a}_j + [s_{k,j,1}, s_{k,j,2}, \ldots, s_{k,j,R_j}]),
\end{align}
where the $s_{k,j,r}$'s are as previously defined

If $\gamma_j = 0$, then the conditional for $\Phi_{k,j}$ is just the prior, $p(\Phi_{k,j})$, as provided in Equation \eqref{prior1}.  [But note that, in practice, we would never need to sample from the conditional in this case].

\paragraph{Inferring $\Phi_{0,j}$}
As we did for $\Phi_{k,j}$, we adopt a Dirichlet prior for 
$\Phi_{0,j}$, 
\begin{equation}
\Phi_{0,j} \sim \mbox{Dirichlet}({\bf a}_j^{(0)}),\label{prior0}
\end{equation}
where ${\bf a}_j^{(0)} = [a_{j,1}^{(0)}, \ldots, a_{j,R_j}^{(0)}]$ is the vector of concentration parameters.  
It is clear from Equation \eqref{fsjoint} that the conditional for $\Phi_{0,j}$ is
\begin{equation}
p(\Phi_{0,j}|{\bf x}_1,\ldots, {\bf x}_n,z_1, \ldots, z_n, \boldsymbol{\theta}, \boldsymbol{\Phi}_{-{k,j}}, \boldsymbol{\Phi}_0, \pi, \alpha, \gamma) \propto p(\Phi_{0,j}) \prod_{i=1}^n  \prod_{j=1}^J  \phi_{0,j,x_{ij}}^{\mathbb{I}(\gamma_j=1)}.
\end{equation}

If $\gamma_j = 1$, then the conditional for $\Phi_{0,j}$ is just the prior, $p(\Phi_{0,j})$, as provided in Equation \eqref{prior0}.  [But note that, in practice, we would never need to sample from the conditional in this case].    

If $\gamma_j = 0$, then the conditional for $\Phi_{0,j}$ is the posterior for $\Phi_{0,j}$ given {\em all} observations, i.e. 
\begin{align}
\Phi_{0,j}|\gamma_j = 1,{ x}_{1,j}, { x}_{2,j}, \ldots, { x}_{n,j}, {\bf a}_j^{(0)} \sim \mbox{Dirichlet}({\bf a}_j^{(0)} + [s_{j,1}, s_{j,2}, \ldots, s_{j,R_j}]),
\end{align}
where $s_{j,r}$ is defined to be the number of observations for which the $j$-th variable is in category $r$.

\subsubsection{Conditional for $z_i$ }

Let us temporarily return to the case of finite $K$, and recall Equation \eqref{finite_nomarg}, which provides the conditional for $z_i$ (assuming that the $\theta_k$'s are sampled, rather than integrated out).  Introducing the $\gamma_j$'s (to allow for variable selection), and exploiting the conditional independence of the variables given $z_i$, Equation \eqref{finite_nomarg} becomes: 
\begin{align}
p(z_i = k| {\bf x}_1,\ldots, {\bf x}_n,\boldsymbol{\theta}, z_{-i}, \alpha, \gamma) &= \frac{1}{Z} \frac{n_{-i,k} + \alpha/K}{n - 1 + \alpha}  f_{ y}({ y}_i | \theta_k)\left(\prod_{j: {\gamma_j} = 1}f_{\bf x}(x_{ij} | \Phi_{k,j})\right)\left(\prod_{j: {\gamma_j} = 0}f_{\bf x}(x_{ij} | \Phi_{0,j})\right)\\
&= \frac{1}{Z} \frac{n_{-i,k} + \alpha/K}{n - 1 + \alpha}  \theta_{k,y_i}\left(\prod_{j: {\gamma_j} = 1} \phi_{k,j,x_{ij}}\right)\left(\prod_{j: {\gamma_j} = 0}\phi_{0,j,x_{ij}}\right).
\label{finite_nomarg_fs}
\end{align} 

\paragraph{Marginalising $\theta_k$}
After some algebra, it is straightforward to show that marginalising the component-specific parameters, $\theta_k$, gives the following for the conditional for~$z_i$:
\begin{align}
p(z_i = k| {\bf x}_1,\ldots, {\bf x}_n,z_{-i}, \alpha) &= \frac{1}{Z} \frac{n_{-i,k} + \alpha/K}{n - 1 + \alpha} \times \int_{\theta_k}f_{y}(y_i | \theta_k) p(\theta_k|\{{ y}_{m}\}_{m\in \mathcal{M}_{-i,k}}) d\theta_k\notag\\
&\times \prod_{j: {\gamma_j} = 1} \left(\int_{\Phi_{k,j}}f_{{\bf x}}(x_{ij} | \Phi_{k,j}) p(\Phi_{k,j}|\{{ x}_{mj}\}_{m\in \mathcal{M}_{-i,k}}) d\Phi_{k,j}\right)\notag\\
&\times \prod_{j: {\gamma_j} = 0} \left(\int_{\Phi_{0,j}}f_{{\bf x}}(x_{ij} | \Phi_{0,j}) p(\Phi_{0,j}|\{{ x}_{m}\}_{m\in \mathcal{M}_{-i}}) d\Phi_{0,j}\right),\label{condci4_fs}
\end{align} 
where 
\begin{itemize}
\item $\mathcal{M}_{-i} = \{m : m \in \{1, \ldots, n \} \mbox{ and } m \ne i \}$; and 
\item $\mathcal{M}_{-i,k} = \{m : m \in \{1, \ldots, n \}\mbox{ and }  m \ne i\mbox{ and }  z_m = k \}$.  
\end{itemize}
Thus, as previously, $p(\theta_k|\{{ y}_{m}\}_{m\in \mathcal{M}_{-i,k}})$ is the posterior for $\theta_k$ given all observations (excluding $y_i$) currently associated with the $k$-th component (and similarly for $\Phi_{k,j}$).  Also as before, if the $k$-th component is {\em empty} we define $p(\theta_k|\{{ y}_{m}\}_{m\in \mathcal{M}_{-i,k}}) := p(\theta_k)$ to be the prior for $\theta_k$ (and similarly for $\Phi_{k,j}$).  

Note that the expression involving $\Phi_{0,j}$ is slightly different: here, we have the posterior for $\Phi_{0,j}$ given {\em all} observations except the $i$-th.  Since this expression does not involve $k$, we may absorb this term into the normalising constant, $Z$, so that:   

\begin{align}
p(z_i = k| {\bf x}_1,\ldots, {\bf x}_n,z_{-i}, \alpha) &= \frac{1}{Z'} \frac{n_{-i,k} + \alpha/K}{n - 1 + \alpha} \times \int_{\theta_k}f_{y}(y_i | \theta_k) p(\theta_k|\{{ y}_{m}\}_{m\in \mathcal{M}_{-i,k}}) d\theta_k\notag\\
&\times \prod_{j: {\gamma_j} = 1} \left(\int_{\Phi_{k,j}}f_{{\bf x}}(x_{ij} | \Phi_{k,j}) p(\Phi_{k,j}|\{{ x}_{mj}\}_{m\in \mathcal{M}_{-i,k}}) d\Phi_{k,j}\right).\label{condci5_fs}
\end{align} 
\paragraph{Conditional for $z_i$ ($K$ infinite)}

Proceeding as previously, we consider the limit as $K \rightarrow \infty$.  In this case, the conditionals for $z_i$ are as follows:  

\noindent If $k = z_j$ for some $j \ne i$, then\qquad
\begin{align}
p(z_i=k|{\bf x}_1,\ldots, {\bf x}_n,z_{-i}, \alpha) &= b\frac{n_{-i,k}}{n - 1 + \alpha} \times \frac{\int_{\theta_k}f_{y}(y_i,\{{ y}_{m}\}_{m\in \mathcal{M}_{-i,k}} | \theta_k) p(\theta_k) d\theta_k}{\int_{\theta_k}f_{y}(\{{ y}_{m}\}_{m\in \mathcal{M}_{-i,k}} | \theta_k) p(\theta_k) d\theta_k}     \notag\\ 
&\times \frac{\prod_{j: {\gamma_j} = 1} \left(\int_{\Phi_{k,j}}f_{{\bf x}}(x_{ij},\{{ x}_{mj}\}_{m\in \mathcal{M}_{-i,k}} | \Phi_{k,j}) p(\Phi_{k,j}) d\Phi_{k,j}\right)}{\prod_{j: {\gamma_j} = 1} \left(\int_{\Phi_{k,j}}f_{{\bf x}}(\{{ x}_{mj}\}_{m\in \mathcal{M}_{-i,k}} | \Phi_{k,j}) p(\Phi_{k,j}) d\Phi_{k,j}\right)}. \label{infcond1fs}
\end{align}
Moreover, \qquad
\begin{align}
p(z_i \ne z_j \mbox{ for all } j \ne i| {\bf x}_1,\ldots, {\bf x}_n, z_{-i}, \alpha) &= b\frac{\alpha}{n - 1 + \alpha} \times \int_{\theta_k}f_{y}(y_i | \theta_k) p(\theta_k) d\theta_k \notag \\&\times  \prod_{j: {\gamma_j} = 1} \left(\int_{\Phi_{k,j}}f_{{\bf x}}(x_{ij} | \Phi_{k,j}) p(\Phi_{k,j}) d\Phi_{k,j}\right),\label{infcond2fs}
\end{align}
where $b$ is a normalising constant that ensures that 
\begin{equation}
p(z_i \ne z_j \mbox{ for all } j \ne i| {\bf x}_1,\ldots, {\bf x}_n, z_{-i}, \alpha) + \sum_{k \in \mathcal{C}_{-i}} p(z_i=k|{\bf x}_1,\ldots, {\bf x}_n,z_{-i}, \alpha) = 1. \notag
\end{equation}

Note that these expressions are essentially the same as those provided in Equations \eqref{infcond1} and \eqref{infcond2}, {\em except} that the irrelevant variables do not contribute to the conditionals (as we would hope).  The marginal likelihoods may be evaluated as previously described (see Sections \ref{marginphi} and \ref{margintheta}). 

\subsubsection{Sampling $\gamma_j$}
By examination of Equation \eqref{fsjoint}, the conditional for $\gamma_j$ is immediately
\begin{equation}
p(\gamma_j|{\bf x}_1,\ldots, {\bf x}_n,z_1, \ldots, z_n, \boldsymbol{\theta}, \boldsymbol{\Phi}, \boldsymbol{\Phi}_0, \pi, \alpha, \gamma_{-j}) \propto p(\gamma_j)\left(\prod_{i=1}^n   \phi_{z_i,j,x_{ij}}^{\mathbb{I}(\gamma_j=1)}\phi_{0,j,x_{ij}}^{\mathbb{I}(\gamma_j=0)}\right).\label{gammacond1}
\end{equation}

Since $\gamma_j \in \{0,1\}$, we have:
\begin{equation}
p(\gamma_j = 1|{\bf x}_1,\ldots, {\bf x}_n,z_1, \ldots, z_n, \boldsymbol{\theta}, \boldsymbol{\Phi}, \boldsymbol{\Phi}_0, \pi, \alpha, \gamma_{-j}) = \frac{1}{z} p(\gamma_j = 1)\prod_{i=1}^n   \phi_{z_i,j,x_{ij}},\label{gammacond2a}
\end{equation}
and
\begin{equation}
p(\gamma_j = 0|{\bf x}_1,\ldots, {\bf x}_n,z_1, \ldots, z_n, \boldsymbol{\theta}, \boldsymbol{\Phi}, \boldsymbol{\Phi}_0, \pi, \alpha, \gamma_{-j}) = \frac{1}{z}p(\gamma_j = 0)\prod_{i=1}^n   \phi_{0,j,x_{ij}},\label{gammacond2b}
\end{equation}
where $$z = p(\gamma_j = 1)\prod_{i=1}^n   \phi_{z_i,j,x_{ij}} + p(\gamma_j = 0)\prod_{i=1}^n   \phi_{0,j,x_{ij}}.$$
\paragraph{Marginalising $\Phi_{k,j}$ and $\Phi_{0,j}$}
As previously, we may also marginalise the $\Phi_{k,j}$ and $\Phi_{0,j}$ parameters.  Define $X_{k} = \{ {\bf x}_{i_1}, {\bf x}_{i_2}, \ldots, {\bf x}_{i_{n_k}} \}$ to be the set of observations associated with component $k$, and $X_{k, j} = \{ { x}_{i_1,j}, { x}_{i_2,j}, \ldots, { x}_{i_{n_k},j} \}$ to be the set containing the $j$-th elements of the vectors in $X_{k}$ (as before).  Also define $\mathcal{C} = \{k:  z_i = k \mbox{ for some } i \}$ to be the set of all current component labels, and $X_{\cdot, j} = \cup_{k \in \mathcal{C}} \{ X_{k, j}  \}$ to be the set containing the $j$-th elements of all observations.

We then obtain 
\begin{equation}
p(\gamma_j = 1|{\bf x}_1,\ldots, {\bf x}_n,z_1, \ldots, z_n, \boldsymbol{\theta}, \pi, \alpha, \gamma_{-j}) = \frac{1}{z'} p(\gamma_j = 1)\prod_{k\in \mathcal{C}}  p(X_{k,j}|{\bf a}_j),\label{gammacond3a}
\end{equation}
and
\begin{equation}
p(\gamma_j = 0|{\bf x}_1,\ldots, {\bf x}_n,z_1, \ldots, z_n, \boldsymbol{\theta}, \pi, \alpha, \gamma_{-j}) = \frac{1}{z'}p(\gamma_j = 0)p(X_{\cdot,j}|{\bf a}_j^{(0)}),\label{gammacond3b}
\end{equation}
where $z'$ is a normalising constant that ensures that the probabilities sum to 1.  The marginal likelihoods may be evaluated as previously described (see Section \ref{marginphi}).  

\subsection{Multi-view clustering}\label{multiview}
The method for variable selection described in the previous section may be considered to be a special case of {\em multi-view clustering}, in which we allow the data to possess multiple clustering structures (depending upon which group of variables we consider to be ``relevant").  In the variable selection case, we model the data as possessing 2 clustering structures: one non-trivial clustering structure (to which the ``relevant" variables contribute), and one trivial clustering structure (comprising just a single cluster, to which the ``irrelevant" variables contribute).  This may be straightforwardly extended to $L$ clustering structures, by allowing $\gamma_j$ to be a categorical variable whose value indicates the clustering structure to which the $j$-th variable contributes. 

\subsubsection{Notational conventions}     
We assume that $\gamma_j \in \{0, 1, 2, \ldots, V-1\}$ is a categorical variable with $L$ categories.  As previously, we assume that if $\gamma_j=0$ then the $j$-th variable is completely irrelevant, and does not contribute to any clustering structure (except the trivial structure comprising one cluster).  

\begin{align*}
f({\bf x}_i, y_i | \cdots) &= \left(\prod_{j=1}^J\left(\prod_{v=1}^{V-1} f_{\bf x}( x_{ij} | \theta^{(v)}_{z_i^{(v)}})^{\mathbb{I}(\gamma_j = v)}\right)f_{\bf x}( x_{ij} | \theta_{0})^{\mathbb{I}(\gamma_j = 0)}\right)f_{\bf y}({ y}_i | \phi_{z_i^{(1)}}, {\bf w}_i, \beta)\\
\end{align*}

\begin{align*}
p(\gamma_j = 0 | \cdots) &\propto  p(\gamma_j = 0) \prod_{i=1}^n  f_{\bf x}( x_{ij} | \theta_{0})\\
p(\gamma_j = v | \cdots) &\propto  p(\gamma_j = v) \prod_{i=1}^n  f_{\bf x}( x_{ij} | \theta_{z_i^{(v)}}^{(v)}) \mbox{\qquad \qquad for $v = 1, \ldots, V-1$}\\
\end{align*}

For $\ell = 1, \ldots, L-1$, we assume that there are non-overlapping variable sets $\mathcal{S}_\ell$, such that each $\mathcal{S}_\ell$ possesses its own clustering structure.  Within each $\mathcal{S}_\ell$, we model the data using a mixture model.  To this end, we introduce latent component allocation variables, $z_i^{(\ell)}$, which indicate the component in the model for $\mathcal{S}_\ell$ that is responsible for generating the $i$-th observation.  Similarly, we define $\Phi_k^{(\ell)}$ to be the parameters associated with the $k$-th component in the $\ell$-th mixture model.   

We moreover define $\gamma_j = \ell$ if the $j$-th variable is in $\mathcal{S}_\ell$.   Note that, a priori, we do not know which variables belong to which $\mathcal{S}_\ell$, so we must perform inference for the $\gamma_j$'s.

Finally, we introduce another categorical variable, $\nu \in  \{1, 2, \ldots, L-1\}$, such that the clustering structure present in $\mathcal{S}_\nu$ is the one that is relevant for profile regression, i.e. the likelihood for $y_i$, 
$$f_{\bf y}(y_i|\boldsymbol{\theta}, z_i^{(1)}, \ldots, z_i^{(L-1)}, \nu) = f_{\bf y}(y|\boldsymbol{\theta}_{z_i^{(\ell)}}),$$      
depends only on the $z_i^{(\ell)}$ component allocation variable. 

Our conditional model for ${\bf x}_i = [{\bf x}_i, y_i]^\top$ (given the $z_i^{(\ell)}$'s, $\gamma_j$'s, and $\nu$) is then
\begin{equation}
\begin{split}
p({\bf x}_i|z_i^{(1)}, \ldots, z_i^{(L-1)}, &\boldsymbol{\theta}, \Phi_0, \Phi^{(1)}, \ldots, \Phi^{(L-1)}, \gamma, \nu) \\
&=  f_{\bf y}(y|\boldsymbol{\theta}_{z_i^{(\ell)}})\prod_{j=1}^J\left( f(x_{ij}|\phi_{0, j, x_{ij}})^{\mathbb{I}(\gamma_j=0)}\prod_{\ell = 1}^{L-1}\left(f(x_{ij}|\phi^{(\ell)}_{z_i^{(\ell)}, j, x_{ij}})^{\mathbb{I}(\gamma_j=\ell)}\right)\right).
\end{split}
\end{equation}

\subsubsection{Conditionals for Gibbs sampling}
We may follow the same method of argument presented in Section 3 in order to derive the required conditionals.  For brevity, we provide the conditionals obtained after integrating out the component-specific parameters.  
\paragraph{Conditionals for $z_i^{(\ell)}$: $\nu = \ell$ case}
In this case, we are dealing with the clustering structure that is relevant for profile regression, so must include the contribution of the responses variable, $y$.  We have (cf. Equations \eqref{infcond1fs} and \eqref{infcond2fs}):
\noindent If $k = z_j^{(\ell)}$ for some $j \ne i$, then\qquad
\begin{align}
p(z_i^{(\ell)}=k|{\bf x}_1,\ldots, {\bf x}_n,z_{-i}^{(\ell)}, \alpha^{(\ell)}, \nu = \ell) &= b^{(\ell)}\frac{n_{-i,k}^{(\ell)}}{n - 1 + \alpha^{(\ell)}} \times \frac{\int_{\theta_k}f_{y}(y_i,\{{ y}_{m}\}_{m\in \mathcal{M}^{(\ell)}_{-i,k}} | \theta_k) p(\theta_k) d\theta_k}{\int_{\theta_k}f_{y}(\{{ y}_{m}\}_{m\in \mathcal{M}^{(\ell)}_{-i,k}} | \theta_k) p(\theta_k) d\theta_k}     \notag\\ 
&\times \frac{\prod_{j: {\gamma_j} = 1} \left(\int_{\Phi^{(\ell)}_{k,j}}f_{{\bf x}}(x_{ij},\{{ x}_{mj}\}_{m\in \mathcal{M}^{(\ell)}_{-i,k}} | \Phi^{(\ell)}_{k,j}) p(\Phi^{(\ell)}_{k,j}) d\Phi^{(\ell)}_{k,j}\right)}{\prod_{j: {\gamma_j} = 1} \left(\int_{\Phi^{(\ell)}_{k,j}}f_{{\bf x}}(\{{ x}_{mj}\}_{m\in \mathcal{M}^{(\ell)}_{-i,k}} | \Phi^{(\ell)}_{k,j}) p(\Phi^{(\ell)}_{k,j}) d\Phi^{(\ell)}_{k,j}\right)}, \label{infcond1mc1}
\end{align}
where $n_{-i,k}^{(\ell)}$ is the number of observations, ${\bf x}_m$, for which $z_m^{(\ell)} = k$ (excluding the $i$-th observation), and $\mathcal{M}_{-i,k}^{(\ell)} = \{m : m \in \{1, \ldots, n \}\mbox{ and }  m \ne i\mbox{ and }  z_m^{(\ell)} = k \}$.

\noindent Moreover, \qquad
\begin{align}
p(z_i^{(\ell)} \ne z_j^{(\ell)} \mbox{ for all } j \ne i| {\bf x}_1,\ldots, {\bf x}_n, c^{(\ell)}_{-i}, \alpha^{(\ell)}, \nu = \ell) &= b^{(\ell)}\frac{\alpha^{(\ell)}}{n - 1 + \alpha^{(\ell)}} \times \int_{\theta_k}f_{y}(y_i | \theta_k) p(\theta_k) d\theta_k \notag \\&\times  \prod_{j: {\gamma_j} = 1} \left(\int_{\Phi^{(\ell)}_{k,j}}f_{{\bf x}}(x_{ij} | \Phi^{(\ell)}_{k,j}) p(\Phi^{(\ell)}_{k,j}) d\Phi^{(\ell)}_{k,j}\right),\label{infcond2mc1}
\end{align}
where $b^{(\ell)}$ is a constant that ensures that the probabilities sum to 1, and $\alpha^{(\ell)}$ is the concentration parameter of the Dirichlet process prior on the mixture weights for the $\nu$-th mixture model.  

\paragraph{Conditionals for $z_i^{(\ell)}$: $\nu \ne \ell$ case }
In this case, the response variable, $y$, does not contribute to the conditionals.  We have (cf. Equations \eqref{infcond1fs} and \eqref{infcond2fs}):
\noindent If $k = z_j^{(\ell)}$ for some $j \ne i$, then\qquad
\begin{align}
p(z_i^{(\ell)}=k|{\bf x}_1,\ldots, {\bf x}_n,z_{-i}^{(\ell)}, \alpha^{(\ell)}, \nu \ne \ell) &= b^{(\ell)}\frac{n_{-i,k}^{(\ell)}}{n - 1 + \alpha^{(\ell)}}  \frac{\prod_{j: {\gamma_j} = \ell} \left(\int_{\Phi^{(\ell)}_{k,j}}f_{{\bf x}}(x_{ij},\{{ x}_{mj}\}_{m\in \mathcal{M}_{-i,k}} | \Phi^{(\ell)}_{k,j}) p(\Phi^{(\ell)}_{k,j}) d\Phi^{(\ell)}_{k,j}\right)}{\prod_{j: {\gamma_j} = \ell} \left(\int_{\Phi^{(\ell)}_{k,j}}f_{{\bf x}}(\{{ x}_{mj}\}_{m\in \mathcal{M}_{-i,k}} | \Phi^{(\ell)}_{k,j}) p(\Phi^{(\ell)}_{k,j}) d\Phi^{(\ell)}_{k,j}\right)}. \label{infcond1mc1}
\end{align}
where $n_{-i,k}^{(\ell)}$ is the number of observations, ${\bf x}_m$, for which $z_m^{(\ell)} = k$ (excluding the $i$-th observation).
Moreover, \qquad
\begin{align}
p(z_i^{(\ell)} \ne z_j^{(\ell)} \mbox{ for all } j \ne i| {\bf x}_1,\ldots, {\bf x}_n, c^{(\ell)}_{-i}, \alpha^{(\ell)}, \nu \ne \ell) &= b^{(\ell)}\frac{\alpha^{(\ell)}}{n - 1 + \alpha^{(\ell)}} \prod_{j: {\gamma_j} = \ell} \left(\int_{\Phi^{(\ell)}_{k,j}}f_{{\bf x}}(x_{ij} | \Phi^{(\ell)}_{k,j}) p(\Phi^{(\ell)}_{k,j}) d\Phi^{(\ell)}_{k,j}\right),\label{infcond2mc1}
\end{align}
where $b^{(\ell)}$ is a constant that ensures that the probabilities sum to 1, and $\alpha^{(\ell)}$ is the concentration parameter of the Dirichlet process prior on the mixture weights for the $\ell$-the mixture model.  

\subsubsection{Conditional for $\gamma_j$}
Define $X_{k}^{(\ell)} = \{ {\bf x}_{i_1}, {\bf x}_{i_2}, \ldots, {\bf x}_{i_{n_k^{(\ell)}}} \}$ to be the set of observations associated with component $k$ of the $\ell$-th mixture model, and $X_{k, j}^{(\ell)} = \{ { x}_{i_1,j}, { x}_{i_2,j}, \ldots, { x}_{i_{n_k^{(\ell)}},j} \}$ to be the set containing the $j$-th elements of the vectors in $X_{k}^{(\ell)}$.  Also define $\mathcal{C}^{(\ell)} = \{k:  z_i^{(\ell)} = k \mbox{ for some } i \}$ to be the set of all current component labels for the $\ell$-th mixture model.

We have 
\begin{equation}
p(\gamma_j = 0|\ldots) = \frac{1}{z'}p(\gamma_j = 0)p(X_{\cdot,j}|{\bf a}_j^{(0)}),\label{gammacond4b}
\end{equation}
and, for all other $\ell \ne 0$,
\begin{equation}
p(\gamma_j = \ell | \ldots) = \frac{1}{z'} p(\gamma_j = \ell)\prod_{k\in \mathcal{C}^{(\ell)}}  p(X_{k,j}^{(\ell)}|{\bf a}_j^{(\ell)}),\label{gammacond4a}
\end{equation}
where $z'$ is a normalising constant that ensures that the probabilities sum to 1.  The marginal likelihoods may be evaluated as previously described (see Section \ref{marginphi}).  

\subsubsection{Conditional for $\nu$}
The conditional for $\nu$ is rather similar to the conditional for $\gamma_j$.  Define $Y_{k}^{(\ell)} = \{ {y}_{i_1}, {y}_{i_2}, \ldots, {y}_{i_{n_k^{(\ell)}}} \}$.  Then, for $\ell \in \{1, \ldots, L-1\}$, we have:
\begin{equation}
p(\nu = \ell | \ldots) = \frac{1}{z''} p(\nu = \ell)\prod_{k\in \mathcal{C}^{(\ell)}}  p(Y_k^{(\ell)}|{\bf a}_y),\label{nucond}
\end{equation}
where $z''$ is a normalising constant to ensure that probabilities sum to 1.

Note that, when sampling $\nu$, we are effectively comparing different models for partitioning the $y$'s, and selecting from amongst these according to probabilities given by Equation \eqref{nucond} above.

%
%
%
%

%
%
%
%
%

\bibliographystyle{humannat}
\bibliography{biblio}

%
%

\end{document}